\documentclass{article}

\usepackage{graphicx}
\usepackage{amsmath}
\usepackage{amsfonts}
\usepackage{amssymb}
\usepackage{amsthm}
\usepackage{url}
\usepackage{cite} % to list multiple citations in numerical order within the sentence
\usepackage[square,sort,comma,numbers]{natbib}
\usepackage[T1]{fontenc}
\usepackage{hyperref}

\theoremstyle{definition}
\newtheorem{theorem}{Theorem}[section]

\newtheorem{algorithm}{Algorithm}[section]

\newcommand{\bigominus}{{\Large\ominus}}

\newcommand{\intextbigominus}{\textrm{\large$\ominus$}}

\begin{document}

\title{A new threshold model based on tropical mathematics reveals network backbones}
\author{
Ebrahim~L.~Patel\footnote{Corresponding author: ebrahim.patel@lis.ac.uk} \\
{\small The London Interdisciplinary School} \\ 
{\small London E1 1EW, UK } \\  
} 

\maketitle

\begin{abstract}
Maxmin-$\omega$ is a new threshold model, where each node in a network waits for the arrival of states from a fraction $\omega$ of neighborhood nodes before processing its own state, and subsequently transmitting it to downstream nodes.  Repeating this sequence of events leads to periodic behavior.  We show that maxmin-$\omega$ reduces to a smaller, simpler, system represented by tropical mathematics, which forges a useful link between the nodal state update times and circuits in the network.  Thus, the behavior of the system can be analyzed directly from the smaller network structure and is computationally faster.  We further show that these reduced networks: (i) are not unique; they are dependent on the initialisation time, (ii) tend towards periodic orbits of networks -- ``attractor networks."  In light of these features, we vary the initial condition and $\omega$ to obtain statistics on types of attractor networks.  The results suggest the case $\omega =0.5$ to give the most stable system.  We propose that the most prominent attractor networks of nodes and edges may be seen as a `backbone' of the original network.  We subsequently provide an algorithm to construct attractor networks, the main result being that they must necessarily contain circuits that are deemed critical and that can be identified without running the maxmin-$\omega$ system.  Finally, we apply this work to the C. elegans neuron network, giving insight into its function and similar networks.  This novel area of application for tropical mathematics adds to its theory, and provides a new way to identify important edges and circuits of a network.   
\\

\end{abstract}

\section{Introduction}
The simplest way to think about discrete dynamics on a network is in synchrony, where all nodes update their states at the same time.  Crucial insight can be gleaned by modeling dynamics in this way, for example in Boolean networks \cite{kauffman1969}.  On the other hand, such models can overlook interesting processes, such as the mutual exchange of information between nodes before an update \cite{ELP2014}.  Thus, we present maxmin-$\omega$, a model of such information exchange that produces asynchronous dynamics on a network.

%Nodes in the maxmin-$\omega$ system update their states upon receiving input from a proportion $\omega$ of neighborhood nodes.  The maxmin-$\omega$ system, therefore, is asynchronous yet deterministic, in contrast to many traditional asynchronous schemes that rely on stochastic processes (such as \cite{schonfisch} and \cite{fates2005}).  In recent presentations of the system, we briefly mentioned maxmin-$\omega$ as being a new threshold model  \cite{ELP2018complexity}.  Here, we examine the model in more detail.

\begin{figure}[!hbtp]
	\centering
	
	\includegraphics[scale=0.5,trim= 0 105 0 65 , clip]{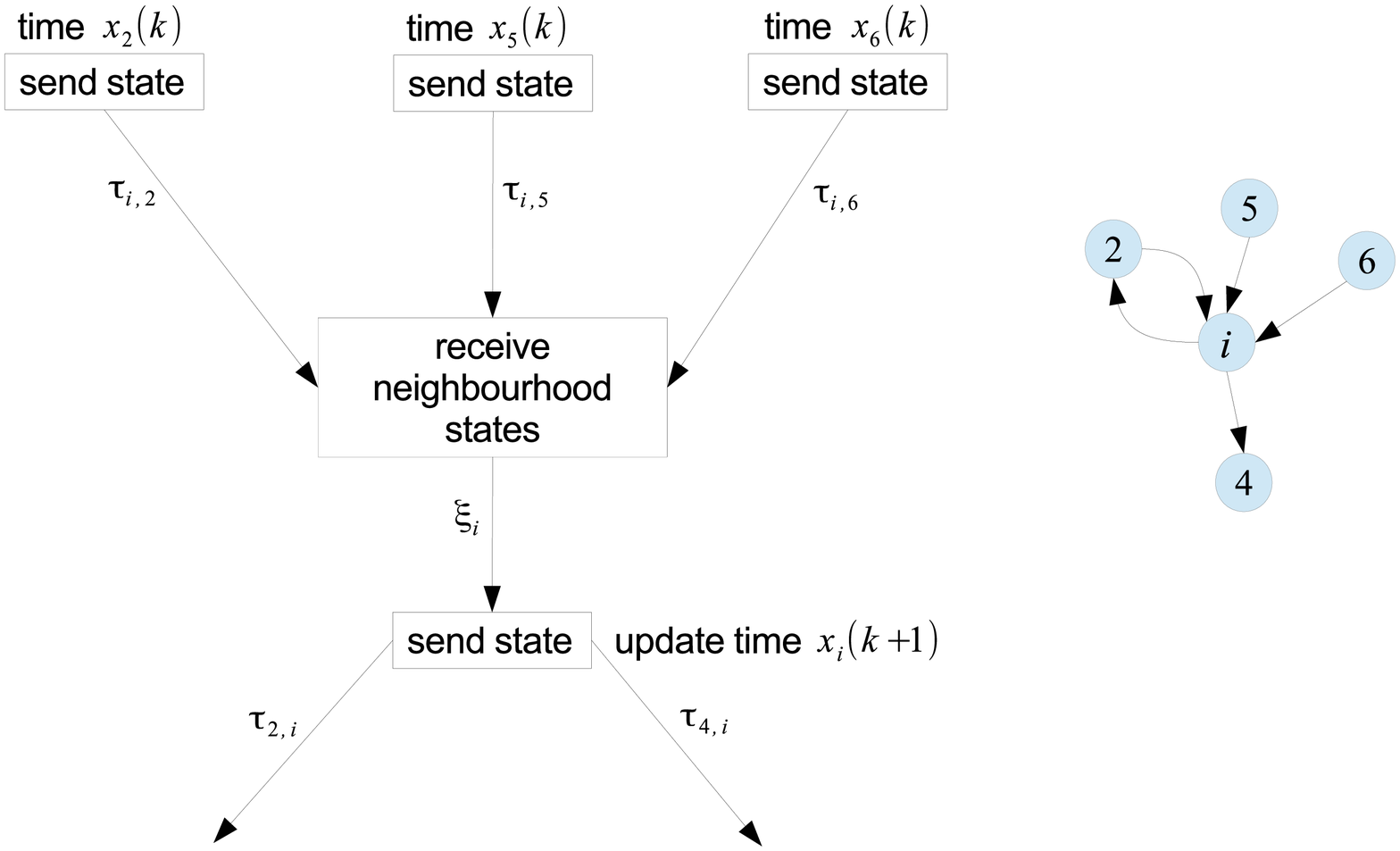}

	\caption{The processes within the $k^\textrm{th}$ cycle that yield a state change for node $i$ due to maxmin-$\omega$.  As an example here, nodes 2, 5, and 6 form the neighborhood of $i$, whilst nodes 2 and 4 are downstream of $i$ (the corresponding network is shown to the right).}
	\label{fig:statechangeExample}
\end{figure}

Each node in the maxmin-$\omega$ system requires knowledge of neighborhood states before updating its own state.  In line with this constraint, maxmin-$\omega$ updates nodal states in epochs that we call cycles, where a \emph{cycle} comprizes the following processes.  First, each neighborhood node $j$ transmits its state to node $i$, which takes some non-zero transmission time $\tau_{ij}$.  Node $i$ waits for a fraction ${\omega\in(0,1]}$ of these arriving neighborhood states before generating its new state, which takes processing time $\xi_i$; this new state is a function of the incoming neighborhood states, though the state dynamics are not our focus here.  Once the node updates its state, it simultaneously transmits the state to downstream nodes, where the cycles are reiterated.  Let $x_i(k+1)$ denote the update time of (the state of) node $i$ in cycle $k+1$ by . The following recurrence relation expresses the above sequence of processes in cycle $k+1$ mathematically.
\begin{equation}\label{equ:maxminOmega}
x_i(k+1) = x_{(\omega)}(k)+\xi_i
\end{equation}
where $x_{(\omega)}(k)$ represents the $k^\textrm{th}$ time of the (last of the) fraction $\omega$ of inputs arriving at $i$.  If there
are $n$ nodes in the neighborhood of $i$, then $x_{(\omega)}(k)$ denotes the arrival time of the $m^\textrm{th}$ input where $m = \lceil\omega n\rceil$.  Figure~\ref{fig:statechangeExample} depicts a cycle of the state change process described above.   
Observe that a cycle $k$ corresponds to the $k^\textrm{th}$ update time, and all nodes update their states at the end of each cycle.  This means that, in a synchronous system (where $\tau_{ij}$ and $\xi_i$ are the same for all edges and nodes respectively), the $k^\textrm{th}$ update time of each node is the same, but in the asynchronous maxmin-$\omega$ system, cycle $k$ for node $i$ does not necessarily occur at the same time as cycle $k$ for node $j$ ($i\neq j$).

These interactions mimic the dynamics of applications such as neuronal networks \cite{mcculloch1943,thul2016} and virus transmission \cite{watts2002,backlund2014}.  Additionally, though the roots are neuroscience and cybernetics, maxmin-$\omega$ fits into the class of \textit{threshold models} that have been further popularised by their application to sociological processes \cite{granovetter1978}.  Thus, the case $m=1$ might model a simple contagion, which is a model of information spread that relies solely on first contact.  For instance, if nodes represent people and edges connect people if they regularly meet each other (let's call them friends), then, at the very least, only one person needs to come into contact with a friend for them to adopt their message.   In contrast, the case $m>1$ might model a complex contagion, where individuals in a social network require social reinforcement to adopt some fad or information supplied by their neighbors \cite{centola2007cascade}.

%Despite the resemblance, three features set this work apart from classical threshold models.  Firstly, since our inspiration is cellular automata, where all nodes process their state for all cycles $k$, we assume all nodes receive input and (if they have at least one outgoing edge) transmit their states.  The system, therefore, is initialized such that each node transmits its state at time $x_i(0)$ ($i=1,\ldots,N$); if there is no outgoing edge, then the node simply updates its state without transmitting.  This is in contrast to threshold models where only one node -- a `seed' node -- initiates some spreading of information.

Despite the resemblance, the maxmin-$\omega$ threshold model is notable for some contrasting features.  Firstly, it is independent of the states of nodes.  Indeed, the resulting nodal state dynamics are interesting in their own right \cite{ELP2016,ELP2018complexity}, but we want to focus on the rich behavior of the underlying asynchronous timing system and maxmin-$\omega$ allows us to do this easily as a separate system to the states.  Nevertheless, in keeping with dynamical systems convention, one is obliged to define a `state'; thus, a maxmin-$\omega$ state is the update time (or firing time) of a node.  A second feature of maxmin-$\omega$ is the element of feedback, i.e., the continuous updating/firing of nodes.  In threshold models such as \cite{granovetter1978} and \cite{schelling1971dynamic}, once the threshold is reached, the cells stop processing, whereas once the threshold $\omega$ is reached, the maxmin-$\omega$ system waits for the next cycle before repeating the same process.  Further developments have been crucial to the advancement of threshold models, but they have not diverged too far from this halting idea \cite{valente1996social,watts2002}.  Recent work such as \cite{oh2018complex} closely resembles our model by incorporating time delays for the uptake of neighborhood information, yet time is also dependent on state in this model such that it halts after a contagion is reached, that is, when a significant proportion of the network first encounters an idea.  These threshold models focus on the states of nodes (e.g., binary) and so are well-suited to social questions like when a population is first `infected';after all, that is more pertinent than knowing if further infections will arise amongst the same group in the future. 

%On the other hand, to model a fuller reality, it seems prudent to study the activity of nodes after their first iterations. In social situations of information diffusion, individuals often have some threshold -- the number of people they are willing to listen to -- beyond which they will relay some information \cite{watts2007influentials,centola2007cascade}.  This kind of verbal communication data is difficult to obtain, but the online world is not.  When exchanging information on social media, individuals tweet, post, and respond regularly; the responses (reactions to other posts, retweets, etc.) have been shown to be influenced by social factors such as the number of followers and followees of a user \cite{petrovic2011rt,suh2010want,zaman2010predicting}.  Although the existense of a threshold-like behavior is yet to be observed, online platforms present a wealth of data to extract, observe, and even hypothetically model this kind of behavior so as to inform other social diffusion applications such as the spreading of illnesses on a social network.  Notwithstanding such epidemic spreading models (as well as neuronal network models mentioned earlier), this type of modern application is a major motivation for maxmin-$\omega$; perhaps it can prove useful for other applications that arise in the future, just like social media was unforeseen in the 1970s when the original threshold models were popularized.  Thus, maxmin-$\omega$ aims to be more realistic due to its asynchronous time delays and modeling capability of recurrent social phenomena.

One might find it helpful to think of maxmin-$\omega$ as an ``iterated threshold model", though it is specifically a discrete event system (DES) \cite{baccelli1992synchronization}.  Traditionally, DESs have been used to optimize scheduling systems, such as those in manufacturing and transport \cite{imaev2008hierarchial,van2006modelling,heidergott}.  There are a few publications that use DES to model threshold properties, though these are often application specific whilst the threshold is highlighted implicitly \cite{chaouiya2006qualitative, chaouiya2011petri}.  Our work explicitly introduces a DES as a threshold model and, moreover, uses it to present a novel way to identify important nodes and edges of a network.   

% However, we have noticed a scarcity of literature that uses DES to model threshold properties; some publications have  .  This seems especially strange when thinking about a particular DES known as a Petri net, which, almost ironically, originated in chemical diffusion \cite{petri1962kommunikationen,petri2008petri}.  Exceptions like \cite{perumalla2012discrete} do exist, though the purpose of the DES in that work is to effectively speed up reaction-diffusion models over networks, and not to necessarily examine and model discrete event systems as a diffusion model in its own right.  Therefore, this is an opportune time to approach threshold phenomena with a new, intuitive, and deterministic model.

In this paper, we exploit tropical mathematics to demonstrate how the maxmin-$\omega$ threshold model on a network allows one to construct smaller, reduced, networks that perform the same function and are computationally more efficient.  We start by introducing tropical mathematics and its link to graph theory in Section~\ref{sec:tropicalgraph}.  We first apply this theory in Section~\ref{sec:reducedminimal}, where nodes update their states upon receipt of the first input; this section acts as a foundation upon which we build in Section~\ref{sec:manyinputs}, where $\omega$ is increased to allow nodes more inputs before updating.  Sections~\ref{sec:manyinputs} and \ref{sec:reducedminimal} go towards informing an algorithm for the construction of the aforementioned reduced networks.  In between, Section~\ref{sec:advantages} highlights the advantages of this work for practitioners, though it should also appeal to current tropical mathematicians as it builds on the theory of critical circuits.  Finally, we apply the model on the C. elegans neuron network in Section~\ref{sec:apps} to demonstrate important functional features of the network that could be gleaned by a maxmin-$\omega$ approach.  We conclude in Section~\ref{sec:conc}.

\section{Tropical mathematics and graph theory}\label{sec:tropicalgraph}

\subsection{The maxmin-$1$ system (``all-inputs")}
We consider the case $\omega=1$ as a first step towards constructing a system of maxmin-$\omega$ equations.  This is equivalent to the condition that nodes wait for all neighborhood information to arrive before updating their states.  First, let us lay some of the groundwork by stating some preliminary definitions that we will use throughout the paper.  Other such definitions will be stated when required.

To every network we associate a \emph{graph}\footnote{We will use the terms ``graph" and ``network" interchangeably, meaning the same thing.}, defined as $\mathcal{G} = (V,E)$, where $E$ is a set of ordered pairs $(a,b)$ (also denoted $ab$) of $V$.  The elements of $V$ are called \emph{vertices} or \emph{nodes}, and those of $E$ are \emph{edges}, where we say the node $a$ is ``upstream of" the node $b$ in the edge $ab$; the node $b$ is ``downstream" of $a$.  In the edge $ab$ we more frequently refer to $a$ as a \textit{neighbour} of $b$.  

Thus, denote the neighborhood of node $i$ by ${\mathcal{N}_i=\{j:j \textrm{ is upstream of }i\}}$.  Then, Equation~(\ref{equ:maxminOmega}) becomes the following.
\begin{equation}\label{equ:maxplus}
x_i(k+1) =\max_{j\in\mathcal{N}_i}\{x_j(k)+\tau_{ij}\}+\xi_i \ .
\end{equation}
Define $\varepsilon=-\infty$ and $e = 0$, and denote by $\mathbb{R}_{\max}$ the set $\mathbb{R}\cup\{\varepsilon\}$. For elements $a, b \in \mathbb{R}_{\max}$, define operations $\oplus$ and $\otimes$ by
$a \oplus b = \max(a, b)$ and $a \otimes b = a + b$.
We refer to the set $\mathbb{R}_{\max}$ together with the operations $\oplus$ and $\otimes$ as \emph{max-plus algebra}, denoted ${\mathcal{R}_{\max} = (\mathbb{R}_{\max}, \oplus, \otimes, \varepsilon, e)}$. 
The quantity $\varepsilon=-\infty$ is the ``zero" (i.e., $\forall x\in\mathbb{R}_{\max}$, $\varepsilon\otimes x = x\otimes\varepsilon = \varepsilon$, and $\varepsilon\oplus x = x\oplus\varepsilon = x$, while $e = 0$ is the ``unit" element (i.e., $\forall x\in\mathbb{R}_{\max}$, $e\otimes x = x\otimes e = x$).

In max-plus notation, (\ref{equ:maxplus}) is now written as
\begin{eqnarray*}
x_i(k+1) &=& \bigoplus_{j\in\mathcal{N}_i}\{\tau_{ij}\otimes x_j(k)\}\otimes \xi_i \\
&=& \bigoplus_{j\in\mathcal{N}_i}\xi_i\tau_{ij}x_j(k)
\end{eqnarray*}
where we omit $\otimes$ for compactness and use the distributivity property of $\otimes$ over $\oplus$.  A system of $N$ equations (for all nodes) is then given in the form
\begin{equation}\label{equ:maxplussystem}
\mathbf{x}(k+1)=P\otimes \mathbf{x}(k)
\end{equation} 
where $\mathbf{x}(k) = (x_1(k),\ldots,x_N(k))^\top$, and the matrix-vector product $P\otimes\mathbf{x}(k)$ is defined as
\begin{equation}
[P\otimes\mathbf{x}(k)]_{i} = \bigoplus_{j=1}^N P_{ij}\otimes x_i(k) = \max_{j=1,\ldots,N} P_{ij} + x_i(k) \ .
\end{equation}
The matrix $P$ itself is defined as
\begin{equation}
P_{ij} = \left\{\begin{array}{rl}
\xi_i\otimes\tau_{ij} & \textrm{if $j$ is a neighbor of $i$} \\
\varepsilon & \textrm{otherwise}
\end{array}\right. \ .
\end{equation}
Since $P$ encodes the connectivity of all nodes, along with weights $\xi_i\otimes\tau_{ij}$ of edges $(j,i)$, it is commonly known as a \emph{weighted adjacency matrix} of the network; since we deal with times in this work, we refer to $P$ as a \emph{timing dependency matrix}.   Sometimes we will denote such a network by $\mathcal{G}(P)$.
%A well-known theory of max-plus algebra says that the asymptotic behavior of this system can be deduced from the connectivity of the network \cite{ELP2014}.  

We continue stating relevant definitions, starting with the sequence ${p = \{e_1,e_2,\ldots,e_n\}}$ of edges.  If there are nodes $v_0,v_1,\ldots,v_n$ (not necessarily distinct) such that $e_j = v_{j-1}v_j$ for $j=1,\ldots,n$, then $p$ is called a walk from $v_0$ to $v_n$.  A walk for which the $e_j$ are distinct is called a path.  Such a path is said to consist of the nodes $v_0,\ldots,v_n$ and to have length $n$, which is denoted $|p|_l = n$.  If $v_n=v_0$ the path is called a circuit, and if the nodes in the circuit are all distinct (i.e., $v_i\neq v_k$ for $i\neq k$), then it is called an elementary circuit.  Finally, define the weight of an edge as a real number associated to that edge, and the 
weight $|p|_w$ of a path $p$ as the sum of weights of all edges constituting the path.  The average weight of $p$ is $|p|_w/|p|_l$; for a circuit, we refer to this quantity as the \emph{average circuit weight}.  

We say that a node $j\in V$ is reachable from node $i\in V$, denoted $i \rightarrow j$, if there exists a path from $i$ to $j$; in this case, we also say that $i$ is a predecessor of $j$.  If either $i=j$ or $i \rightarrow j$ and  $j \rightarrow i$, then we say that node $i$ communicates with node $j$.  (Note that we allow a solitary node $i$ to communicate with itself, even if there is no self-loop $(i,i)$ attached to it.  It is thus possible to partition the node set $V$ of a graph into disjoint subsets $V_i$ such that ${V = V_1\cup V_2 \cup \cdots \cup V_q}$, where each subset $V_i$ contains nodes that communicate with each other but not with other nodes of $V$.  By taking $V_i$ together with edge set $E_i$, each of whose edges has predecessor and successor nodes in $V_i$, we obtain the subgraph $\mathcal{G}_i = (V_i,E_i)$.  We call this subgraph a \emph{maximal strongly connected subgraph} (MSCS) of the original graph $\mathcal{G}=(V,E)$.  If all nodes in a network communicate with each other, then we say the network is strongly connected; in this case, the corresponding timing dependency matrix $P$ is said to be \emph{irreducible}.

These concepts will be used again later, along with the following important result: the asymptotic growth rate of the update times of a max-plus system is governed by the largest average circuit weights of the MSCS's of the network \cite{heidergott}.  More specifically, the update times of each node in a MSCS will share the same asymptotic growth rate. Moreover, let $\lambda_q$ denote the asymptotic growth rate of (the update times of nodes in) a maximal strongly connected subgraph $\textrm{MSCS}_q$ if $\textrm{MSCS}_q$ was considered as a separate graph.  Then the asymptotic growth rate of (the update times of nodes in) both $\textrm{MSCS}_p$ will be equal to 
\begin{equation}\label{equ:asympgrowthratemaxplus}
\bigoplus_{\textrm{MSCS}_q \in \pi(\textrm{MSCS}_p)} \lambda_q 
\end{equation}
where $\pi(\textrm{MSCS}_p)$ denotes all maximal strongly connected subgraphs that are predecessors of $\textrm{MSCS}_q$.  That is, this asymptotic growth rate of a node will be equal to the largest average circuit weight from all MSCS's that are its predecessors.  The corresponding circuits are defined as \emph{critical circuits}; in a transport application, these circuits may be regarded as the bottlenecks that hold up the transport network \cite{baccelli1992synchronization}.  

As a final definition of this section, denote the \emph{cyclicity} of a graph $\mathcal{G}$ by $\sigma_{\mathcal{G}}$.
\begin{itemize}
\item If $\mathcal{G}$ is strongly connected, then $\sigma_{\mathcal{G}}$ equals the greatest common divisor of the lengths of all elementary circuits in $\mathcal{G}$.  If $\mathcal{G}$ consists of only one node without a self-loop, then $\sigma_{\mathcal{G}}$ is defined to be one.
\item If $\mathcal{G}$ is not strongly connected, then $\sigma_{\mathcal{G}}$ equals the least common multiple of the cyclicities of all maximal strongly connected subgraphs of $\mathcal{G}$.
\end{itemize}    

\begin{figure}[!hbp]
\centering
  \includegraphics[scale=0.35,trim=0 140 100 100,clip]{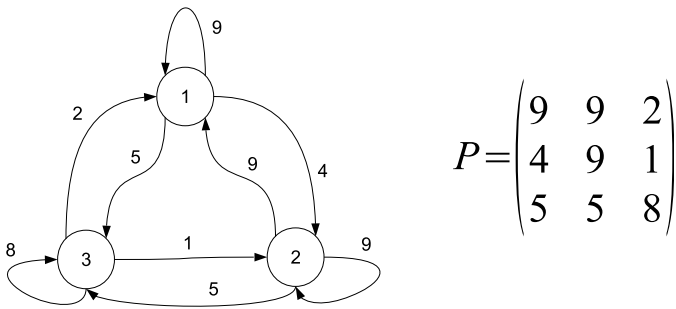}
  \caption{Network on 3 nodes, along with its timing dependency matrix}\label{fig:regular3nbhd3node}
\end{figure}

As an instructive example of some of these concepts, consider the network in Figure~\ref{fig:regular3nbhd3node}, where the weight on edge $(j,i)$ is equal to $\xi_i + \tau_{ij}$.  The timing dependency matrix $P$ of this network is also displayed.
Taking the initial condition $\mathbf{x}(0) = (0,0,0)^\top$ and iterating the max-plus system (\ref{equ:maxplussystem}) a few times, we obtain the following sequence $\mathbf{x}(0),\mathbf{x}(1),\mathbf{x}(2),\ldots$ of update times.
\begin{equation}
\left(\begin{array}{c} 0 \\ 0 \\ 0 \end{array} \right), \left(\begin{array}{c} 9 \\ 9 \\ 8 \end{array} \right), \left(\begin{array}{c} 18 \\ 18 \\ 16 \end{array} \right), \left(\begin{array}{c} 27 \\ 27 \\ 24 \end{array} \right), \left(\begin{array}{c} 36 \\ 36 \\ 32 \end{array} \right), \left(\begin{array}{c} 45 \\ 45 \\ 41 \end{array} \right), \left(\begin{array}{c} 54 \\ 54 \\ 50 \end{array} \right), \left(\begin{array}{c} 63 \\ 63 \\ 59 \end{array} \right), \ldots
\end{equation}
Observe that $\mathbf{x}(k+1) = 9\otimes \mathbf{x}(k)$ for $k\geq 4$, where the addition $\otimes$ is done element-wise.  That is, after four iterations, each nodes updates its state every 9 time units; this is what we refer to as the asymptotic growth rate of the update times.     This result also follows the above, particularly (\ref{equ:asympgrowthratemaxplus}), i.e., for such a strongly connected network as that of Figure~\ref{fig:regular3nbhd3node}, the asymptotic growth rate of update times is the same for all nodes, and is equal to the largest average weight of all elementary circuits -- the average weight of the critical circuits.  Thus, the critical circuits here are the self-loops at nodes 1 and 2, both of whose weights are 9. 

\subsection{Min-plus algebra (``first-input")}
\begin{figure}[!hbp]
\centering
  \includegraphics[scale=0.35,trim=0 120 10 115,clip]{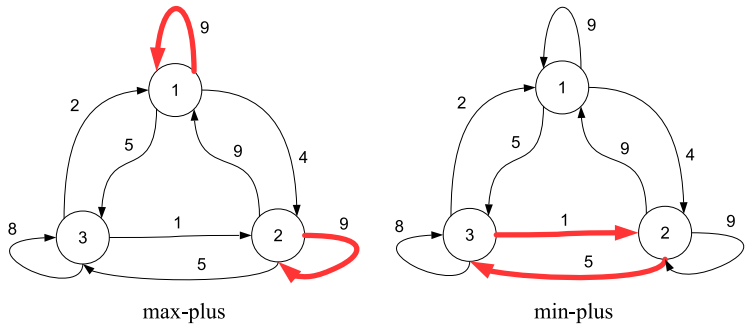}
  \caption{Critical circuits highlighted (as thick, red edges) for two different tropical systems on the network of Figure~\ref{fig:regular3nbhd3node}.  Left: the critical circuits of a max-plus (all-inputs) system, right: the critical circuit of a min-plus (first-input) system.  (Color online.)}\label{fig:criticalcircuits}
\end{figure}

Generally, $m=\lceil\omega|\mathcal{N}_i|\rceil$, so that, when $\omega=1$, $m=|\mathcal{N}_i|$.  The dual to this case is when $m = 1$ (equivalently $\omega = 1/|\mathcal{N}_i|$), such that node $i$ updates upon receiving the first-input.  This gives the following evolution equation for the \emph{first-input} system.
\begin{equation}\label{equ:minplus}
x_i(k+1) =\min_{j\in\mathcal{N}_i}\{x_j(k)+\tau_{ij}\}+\xi_i \ .
\end{equation}
The dual to max-plus algebra is, therefore, min-plus algebra, and the corresponding theory is as one might expect.  For instance, the asymptotic growth rate of the first-input system is governed by the smallest average circuit weights of the maximal strongly connected subgraphs of the network \cite{baccelli1992synchronization}; these are the critical circuits of the first-input system\footnote{The context will clarify whether we are looking at critical circuits of all-inputs or first-input systems.}  For example, the critical circuit of the first-input system on the network of Figure~\ref{fig:regular3nbhd3node} is the circuit of length two between nodes 2 and 3.  Figure~\ref{fig:criticalcircuits} depicts the critical circuits of both of these tropical systems.
We complete this section on min-plus algebra by defining its terms as duals to the max-plus algebraic terms of the previous section.

In keeping with the literature, define $\varepsilon'=+\infty$ and $e = 0$, and denote by $\mathbb{R}_{\min}$ the set $\mathbb{R}\cup\{\varepsilon'\}$. For elements $a, b \in \mathbb{R}_{\min}$, define operations $\ominus$ and $\odot$ by $a \ominus b = \min(a, b)$ and $a \odot b = a + b$.  We refer to the set $\mathbb{R}_{\min}$ together with the operations $\ominus$ and $\odot$ as \emph{min-plus algebra}, denoted $\mathcal{R}_{\min} = (\mathbb{R}_{\min}, \ominus, \odot, \varepsilon', e)$ \cite{heidergott}.  The quantity $\varepsilon'=+\infty$ is the ``zero" (i.e., $\forall x\in\mathbb{R}_{\min}$, $\varepsilon'\odot x = x\odot\varepsilon' = \varepsilon'$, and $\varepsilon'\ominus x = x\ominus\varepsilon' = x$, while $e = 0$ is the ``unit" element (i.e., $\forall x\in\mathbb{R}_{\min}$, $e\odot x = x\odot e = x$).

Memorably, $\mathcal{R}_{\min}$ (resp, $\mathcal{R}_{\max}$) is sometimes referred to as \textit{tropical mathematics}, where $\ominus$ (resp. $\oplus$) is the tropical sum and $\odot$ (resp. $\oplus$) is tropical addition.  The word ``tropical" is an homage to one of the founders of min-plus algebra, who was Brazilian\footnote{The Brazilian founder is Imre Simon \cite{simon1988recognizable}}.  When we need to be specific, we use the terms ``max-plus algebra" and ``min-plus algebra", though there is an obvious appeal to the term ``tropical" for wider distribution of the maxmin-$\omega$ model.

In min-plus notation, (\ref{equ:minplus}) is now written as
\begin{equation}
\renewcommand{\arraystretch}{0.7}
x_i(k+1) =\begin{array}{c}
\\[-2pt]
\stackrel{\bigominus}{{\scriptstyle j\in\mathcal{N}_i}} 
\end{array}\xi_i\tau_{ij}x_j(k)
\end{equation}
where we omit $\odot$ for compactness and use the distributivity property of $\odot$ over $\ominus$.  A system of $N$ equations (for all nodes) is then given in the form
\begin{equation}\label{equ:minplusSystem}
\mathbf{x}(k+1)=Q\odot \mathbf{x}(k)
\end{equation} 
where $\mathbf{x}(k) = (x_1(k),\ldots,x_N(k))^\top$, and the matrix-vector product $Q\odot\mathbf{x}(k)$ is defined as
\begin{equation}
\renewcommand{\arraystretch}{0}[Q\odot\mathbf{x}(k)]_{i} =
\begin{array}{c}
\\[-2pt]
\stackrel{\stackrel{\scriptstyle N}{\bigominus}}{\scriptstyle j=1}\end{array}Q_{ij}\odot x_i(k) = \min_{j=1,\ldots,N} Q_{ij} + x_i(k) \ .
\end{equation}
The matrix $Q$ is defined as
\begin{equation}\label{equ:Qdef}
Q_{ij} = \left\{\begin{array}{rl}
\xi_i\odot\tau_{ij} & \textrm{if $j$ is a neighbor of $i$} \\
\varepsilon' & \textrm{otherwise}
\end{array}\right. \ .
\end{equation}
Note that $\otimes$ is the same as $\odot$; the reason for using either notation is for distinguishing the system in use, which is particularly important for matrix multiplication (next).  Note also that $Q$ is the timing dependency matrix in this first-input system.

It can now be seen that iterations of a tropical system correspond to tropical powers of the matrix $Q$ (or $P$).  Generally, tropical matrix multiplication is defined as follows.

Let $A\in\mathbb{R}_{\max}^{n\times l}, B\in\mathbb{R}_{\max}^{l\times m}$.  Then $[A\otimes B]_{ij} \bigoplus_{k=1}^l a_{ik}\otimes b_{kj}$.  Similarly, for $A\in\mathbb{R}_{\min}^{n\times l}, B\in\mathbb{R}_{\min}^{l\times m}$, $[A\odot B]_{ij} {\intextbigominus}_ {k=1}^{l} a_{ik}\odot b_{kj}$.  For a square matrix $A\in\mathbb{R}_{\max}^{N\times N}$, the $k^\textrm{th}$ power of $A$ is defined as 
\begin{equation}\label{equ:powers_maxplusA}
A^{\otimes k} = \underbrace{A\otimes A\otimes \cdots \otimes A}_{k \textrm{ times}} \ .
\end{equation}
For $A\in\mathbb{R}_{\min}^{N\times N}$, we have 
\begin{equation}\label{equ:powers_minplusA}
A^{\odot k} = \underbrace{A\odot A\odot \cdots \odot A}_{k \textrm{ times}} \ .
\end{equation}

We will also require the definition of tropical matrix addition in both algebras.  For $A,B\in\mathbb{R}_{\max}^{n\times m}$, the sum is defined by $[A\oplus B]_{ij} = a_{ij}\oplus b_{ij}$ $(=\max (a_{ij},b_{ij}))$.  For $A,B\in\mathbb{R}_{\min}^{n\times m}$, we have $[A\ominus B]_{ij} = a_{ij}\ominus b_{ij}$ $(=\min (a_{ij},b_{ij}))$.

The operators $\oplus$, $\ominus$, and $\otimes$ (which is the same as $\odot$) linearize our tropical systems (\ref{equ:maxplussystem}) and (\ref{equ:minplusSystem}).  Seen in this new light, the correspondence to conventional mathematics is made clear, making analysis a much simpler task; for example, both tropical systems possess analogous eigenvalues and eigenvectors \cite{baccelli1992synchronization}.

\subsection{Asymptotics}\label{subsec:asymptotics}
In general, define the general maxmin-$\omega$ system as a function $\mathcal{M}:\mathbb{R}^N\rightarrow\mathbb{R}^N$ whose components $\mathcal{M}_i$ are of the form of
Equ.~(\ref{equ:maxminOmega}).  We represent a system of $N$ such equations by the following.
\begin{equation} \label{equ:maxminSystem}
\mathbf{x}(k+1) = \mathcal{M}(\mathbf{x}(k))
\end{equation}
for $k\geq0$, where $\mathbf{x}(k)=(x_1(k),x_2(k),\ldots,x_N(k))$.

Denote by $\mathcal{M}^p(\mathbf{x})$ the action of applying $\mathcal{M}$ to a vector $\mathbf{x}\in\mathbb{R}^N$ a total of $p$ times, i.e.,
$\mathcal{M}^p(\mathbf{x})=\underbrace{
	\mathcal{M}(\mathcal{M}(\cdots(\mathcal{M}
}_{p\mbox{ times}}
(\mathbf{x}))\cdots))$.

	If it exists, the \emph{cycletime vector} of $\mathcal{M}$ is $\chi(\mathcal{M})$ and is defined as $\lim_{k\rightarrow\infty}(\mathcal{M}^k(\mathbf{x})/k)$.	For some $k\geq0$, consider the set of vectors
	\begin{displaymath}
	\mathbf{x}(k),\mathbf{x}(k+1),\mathbf{x}(k+2),\ldots\in\mathbb{R}^N
	\end{displaymath}
	where $\mathbf{x}(n)=\mathcal{M}^n(\mathbf{x}(0))$ for all $n\geq0$.  Given some $x_i(0)\in\mathbb{R}$, the set $x_i(k),x_i(k+1),x_i(k+2),\ldots$ is called a \emph{periodic regime} of $i\in\mathbb{N}$ if there
	exists $\mu_i\in\mathbb{R}$ and finite numbers $\rho_i,K_i\in\mathbb{N}$ such that, for $k\geq K_i$,
	\begin{displaymath}
	x_i(k+\rho_i)=\mu_i+x_i(k) \ .
	\end{displaymath}
	The \emph{period} of the regime is $\rho_i$, and the integer $K_i$ is called
	the \emph{transient time}.
	
Under our initial conditions (namely $x_i(0)$ is finite for all $i$), $K_i$ will be finite and $\chi(\mathcal{M}) = \left(\chi_1,\ldots,\chi_n\right)$ will be unique  (see \cite{heidergott}, Theorem 12.7).  Moreover, it can be shown that the cycletime of node $i$ is equal to $\mu_i/\rho_i$, which is the asymptotic growth rate of update times of $i$.  

Consequently, we can define a critical circuit more generally as follows.  Consider a maxmin-$\omega$ system $\mathcal{M}$ on a network $\mathcal{G}$ that can be decomposed into a set $\{\textrm{MSCS}_p|p=1,\ldots,q\}$ of $q$ maximal strongly connected subgraphs.  Denote by $\lambda_p$ the cycletime of $\textrm{MSCS}_p$ if $\textrm{MSCS}_p$ was a separate graph; this cycletime is the average weight of an elementary circuit in $\textrm{MSCS}_p$, and its value can be found by iterating $\mathcal{M}$ and observing the periodic regime as above.  Then the cycletime of a node contained in $\textrm{MSCS}_{p'}\in \mathcal{G}$ is equal to the cycletime $\lambda_p$ of some $\textrm{MSCS}_p$, where $\textrm{MSCS}_p$ is a predecessor of  $\textrm{MSCS}_{p'}$.  The corresponding elementary circuits are the \emph{critical circuits of the maxmin-$\omega$} system.  

Note that this equivalence between circuits and cycletime holds true for the tropical systems; there, ``an elementary circuit" is a circuit with maximal average weight and the minimal average weight for the all-inputs and first-input systems, respectively.  E.g., see (\ref{equ:asympgrowthratemaxplus}).

\section{Reduced networks: a minimal case}\label{sec:reducedminimal}
The minimal reduced network arises in the first-input system, so this is our focus in this section.  This will help us to make inroads into the general maxmin-$\omega$ system, i.e., for all $\omega$, which we will also refer to as the ``many-inputs" system.  Let us briefly describe a few relevant applications of the first-input system because there is actually much more that one can intuit here.  

\begin{enumerate}
\item \textbf{Shortest paths.} Taking powers of the timing dependency matrix gives us information about paths between nodes.  Specifically, further to the definition of $Q$ in (\ref{equ:Qdef}), let $Q_{ij} = 0$ if $j=i$.  If all the circuits in $\mathcal{G}$ have positive weight, then the entries in $Q^{\odot l}$, the $l^\textrm{th}$ tropical power of $Q$, are such that $Q^{\odot l}_{ii} = 0$ for all $i$ ($i \neq j$).  Moreover, $Q^{\odot l}_{ij}$ is the weight of a minimal weight walk from $v_j$ to $v_i$ containing at most $l$ edges when $\mathcal{G}$ contains such a walk; otherwise $Q^{\odot l}_{ij} = \varepsilon'$ if no such walk exists.  If there are $N$ nodes in a weighted network, it follows that the weight of a minimal-weight path from $v_j$ to $v_i$ is given by $Q^{\odot (N-1)}_{ij}$.   

\item \textbf{Scheduling.} We can, in fact, write Bellman's equations for finding the shortest paths between two nodes in a network as a min-plus algebraic equation (simply replace the $\min$ and plus operators with $\ominus$ and $\odot$ respectively).  This can then be applied in scheduling and planning networks, which are usually acyclic (i.e., contain no circuits).  For each edge in such networks, upstream and downstream nodes respectively represent the start and end of an activity, with the edge weight representing time taken to complete.  Furthermore, max-plus algebra may be better to perform critical path analysis than standard techniques \cite{kobayashi2012tropical}.  Since it is its dual, a straightforward transformation would convert this model to a min-plus algebraic system.  E.g., multiply all edge weights by $-1$ to then study the equivalent problem of looking for paths of minimal weight instead of maximal weight.

\item \textbf{Simple contagion.} A first-input system may be amenable towards the modelling of a simple contagion.  A simple contagion is a model of information or virus spread that relies solely on first contact.  For instance, if nodes represent people and edges connect friends, then, at the very least, the virus requires just one person to come into contact with a friend to transmit it.  The earliest time this can occur is the minimum of contact times with all friends.  Iterating a first-input system such as (\ref{equ:minplusSystem}) would then give the first contact times of the whole social network.  Subsequently, it may be possible to estimate the earliest time -- worst case scenario -- for the infection of the whole network.
\end{enumerate}

\subsection{Reduced network}

Consider 
\begin{equation}
\mathcal{A}_i(k) = \{j\in\mathcal{N}_i:x_j(k)+\tau_{ij} \leq x_{(\omega)}(k)\}
\end{equation}
which is the set of all nodes whose inputs arrive before or at the same time as the $\omega^\textrm{th}$ input at node $i$ (in cycle $k$).  We call $\mathcal{A}_i(k)$ the set of \textit{affecting nodes of $i$}.  This allows us to introduce the concept of a reduced network.

In cycle $k$ ($k\geq 1$), the \textit{reduced network} is the set of affecting nodes $\mathcal{A}_i(k)$ of all nodes $i$, together with the edges that connect affecting nodes $j\in\mathcal{A}_i(k)$ to their affected node $i$.  The reduced network in cycle $0$ is set as the original network.  When we talk about \emph{reducing the system}, this means we have reduced the network as above.  Sometimes, we refer to the \emph{``reduced system"} instead of reduced network; this is when we are thinking of the original maxmin-$\omega$ dynamics with a reduced network (practically, this means running an all-inputs system on the reduced network; see later).

We can draw up a reduced network for each cycle $k$.  In \cite{ELPthesis}, we show that this sequence of reduced networks asymptotically settles onto a fixed set of reduced networks.  Formally, let us denote by $\mathcal{G}_r(k)$ the reduced network in cycle $k$.  Then we obtain the  sequence (or ``orbit") $\mathcal{G}_r(0),\mathcal{G}_r(1),\mathcal{G}_r(2),\ldots$ of reduced networks such that, for some $k\geq0$, there exists $g\in\mathbb{N}$ such that $\mathcal{G}_r(k+g)=\mathcal{G}_r(k)$.   The set ${\mathcal{O}=\{\mathcal{G}_r(k),\mathcal{G}_r(k+1),\cdots,\mathcal{G}_r(k+g)\}}$ is called a periodic orbit of reduced networks.  This set is dependent on the initial update time $\mathbf{x}(0)$ of the maxmin-$\omega$ system.  We will show this later, but this inevitability of periodic behaviour prompts us to interchangeably refer to $\mathcal{O}$ by the shorter \emph{attractor network}; for convenience, we retain this name even if the period is greater than one, so that an ``attractor network" can be a periodic orbit of reduced networks with period 2, for example.  

The above definitions of affecting nodes and reduced networks is applicable for all $\omega$, so consider again the network of Figure~\ref{fig:regular3nbhd3node} and a first-input system such that $\omega =  1/|\mathcal{N}_i|$ for all nodes $i$.  Note that the timing dependency matrix $P$ is equal to $Q$ here since there are no zero elements (i.e., it is a fully connected network).   We compare orbits and attractor networks arising from two different initial conditions.  

When $\mathbf{x}(0) = (8,5,9)^\top$, we obtain the following sequence of the first few update times.
\begin{equation}
\left(\begin{array}{c} 8 \\ 5 \\ 9 \end{array} \right), \left(\begin{array}{c} 11 \\ 10 \\ 10 \end{array} \right), \left(\begin{array}{c} 12 \\ 11 \\ 15 \end{array} \right), \left(\begin{array}{c} 17 \\ 16 \\ 16 \end{array} \right), \left(\begin{array}{c} 18 \\ 17 \\ 21 \end{array} \right), \left(\begin{array}{c} 23 \\ 22 \\ 22 \end{array} \right), \ldots
\end{equation}
We can see that $\mathbf{x}(k+2) = 6\odot\mathbf{x}(k)$ for $k\geq 1$.  That is, after a transient time $K = 1$, the sequence settles onto the periodic regime with period $\rho = 2$, and cycletime $6/2 = 3$, as verified by the critical circuit of length two, connecting nodes 2 and 3 (Figure~\ref{fig:criticalcircuits}).

A different initial condition, $\mathbf{x}(0) = (3,2,4)^\top$, produces the following sequence of update times.
\begin{equation}
\left(\begin{array}{c} 3 \\ 2 \\ 4 \end{array} \right), \left(\begin{array}{c} 6 \\ 5 \\ 7 \end{array} \right), \left(\begin{array}{c} 9 \\ 8 \\ 10 \end{array} \right), \left(\begin{array}{c} 12 \\ 11 \\ 13 \end{array} \right), \left(\begin{array}{c} 15 \\ 14 \\ 16 \end{array} \right), \left(\begin{array}{c} 18 \\ 17 \\ 19 \end{array} \right), \ldots
\end{equation}
and this yields the periodic regime, $\mathbf{x}(k+1) = 3\odot\mathbf{x}(k)$ for $k\geq 0$.  That is, the sequence immediately settles onto the periodic regime with period $\rho = 1$, and cycletime $3$.  

\begin{figure}[!hbtp]
	\centering
	
	\includegraphics[scale=0.25,trim= 30 100 10 70 , clip]{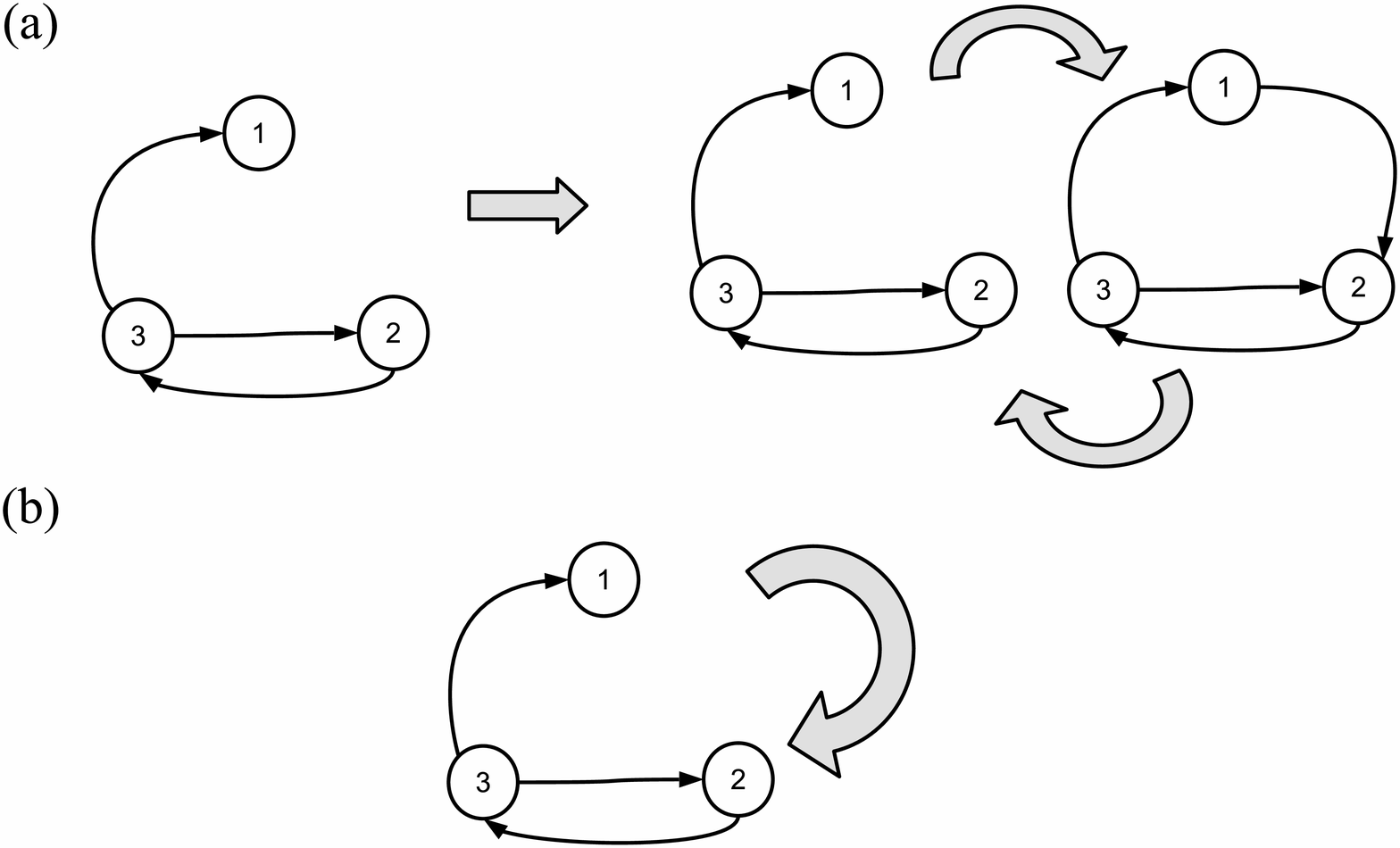} 
	
	\caption{Two different orbits of reduced networks of the first-input system $\mathcal{M}$ with underlying network of Figure~\ref{fig:regular3nbhd3node}.  The first orbit (a) arises from initial condition $\mathbf{x}(0) = (8,5,9)^\top$ and the second orbit (b) is generated by $\mathbf{x}(0)=(3,2,4)^\top$.  Larger arrows indicate the transitions between successive iterations of $\mathcal{M}$.  Edge weights omitted for clarity.}
	\label{fig:reducednetworks_minplus}
\end{figure}

The resulting reduced networks of both sequences are shown in Figure \ref{fig:reducednetworks_minplus}.  Notice the period-2 behaviour is reflected in the first periodic orbit of reduced networks.   

\subsection{Features of the first-input attractor network} \label{subsec:features}
The first feature that we notice is the existence of the critical circuit in both attractor networks in Figure~\ref{fig:reducednetworks_minplus}.  Thus, we conjecture this to be true for all initial conditions.  It should not be difficult to see why.

Firstly, observe that any attractor network $\mathcal{G}_r$ must contain a circuit; this is because every node must have at least one neighbor supplying it with input.  Since the network is finite, we can thus trace the path of these neighbors until we reach a node twice; such a path is a circuit.  Now suppose that $\mathcal{G}_r$ does not contain the critical circuit; in fact, let us generalise this and suppose that $\mathcal{G}_r$ does not contain the critical graph  $\mathcal{G}^{cr}$ (accounting for there being more than one critical circuit).   Then the cycletime of $\mathcal{G}_r$ must be smaller than $\chi$, the cycletime of the first-input system, yielding a contradiction.

To gain further insight into these networks, we computationally collect the different types of attractor networks upon varying $\mathbf{x}(0)$.  From 10000 initial states $x_i(0)$ drawn uniformly at random from 1 to 10 for all $i$, we get that the attractor network (a) of Figure~\ref{fig:reducednetworks_minplus} is obtained 47\% of the time and the attractor network (b) appears in 53\% of output.  As predicted, the critical circuit appears in both of these networks.

The second feature we can interrogate is the period-1 attractor network (b).  Why does this appear when it does?  It is obviously due to the initial condition but notice in Figure~\ref{fig:reducednetworks_minplus} the system instantly settles onto this network, i.e., the transient time is 0.  To better understand this, it first helps to think of cycles as epochs.  When we speak of a reduced network, we specifically mean that reduced network in an epoch $k$.  There are two factors that contribute to an epoch's reduced network: the previous update times $\mathbf{x}(k-1)$, which define the start of the epoch, and the edge weights.  We know the edge weights do not change, so the previous update times must be the sole reason for reduced networks being different for different epochs. Therefore, we need only consider the epoch's previous update times relative to each other. Reduced networks only depend on each node's set of affecting nodes; when this happens is irrelevant. This means that epochs whose vectors $\mathbf{x}(k)$ of update times are parallel will have the same reduced network. 

When reduced networks like Figure~\ref{fig:reducednetworks_minplus}(b) are the same for all epochs $k$, this means the vectors $\mathbf{x}(k)$ are parallel for all $k$.  Successive parallel vectors $\mathbf{x}(k-1)$ and $\mathbf{x}(k)$ are separated by a constant $\lambda$, i.e., $\mathbf{x}(k) = \lambda\odot\mathbf{x}(k-1)$.  Given the min-plus system equation (\ref{equ:minplusSystem}), we can rewrite this as
\begin{equation}
Q\odot \mathbf{x}(k-1) = \lambda\odot\mathbf{x}(k-1)
\end{equation}
which is an eigenvalue-eigenvector equation in min-plus algebra.  That is, $\lambda$ is an eigenvalue of $Q$ and $\mathbf{x}(k-1)$ is an eigenvector.  Since all vectors in our example system are parallel, we infer that the initial condition $\mathbf{x}(0)$ is an eigenvector.

If the system is not initialised to an eigenvector, we may see a transient region before settling onto a periodic regime with period 1.  The first vector $\mathbf{x}(k)$ of this regime must then be an eigenvector.

As in conventional linear algebra, eigenvectors are not unique in min-plus algebra because they are defined up to scalar multiplication.  (It can easily be shown that, if $\mathbf{v}$ and $\mathbf{w}$ are eigenvectors of $A\in\mathbb{R}_{\min}^{n\times n}$ associated
with eigenvalue $\lambda$, then, for $\alpha,\beta\in\mathbb{R}_{\min}$,
$\alpha\odot\mathbf{v}\ominus\beta\odot\mathbf{w}$ is also an
eigenvector of $A$.)

Consider the definition of the \emph{Kleene
star} for any $A\in\mathbb{R}_{\min}^{n\times n}$:
\begin{equation}
A^*\stackrel{\textrm{def}}{=}\bigoplus_{k=0}^{\infty}A^{\otimes k}.
\end{equation}
%For any square matrix $A$, the element $[A^{\otimes k}]_{ij}$ is the
%largest weight for a path of length $k$ from node $j$ to $i$ in the communication graph
%of $A$. Thus, if elements of $A$ are positive, then the elements of
%$A^*$ may tend to infinity. Conversely, if all elements of $A$ are
%nonpositive, then $A^*$ is finite. In particular
It is known that, if circuit weights
in $\mathcal{G}(A)$ are nonpositive, then the Kleene star of a
square matrix over $\mathbb{R}_{\min}$ exists \cite{heidergott}.  The critical graph of $A$, denoted $\mathcal{G}^{cr}(A)=(V^{cr}(A),E^{cr}(A))$, is defined as the graph consisting only of the critical circuits of $\mathcal{G}(A)$.  
\begin{theorem}\label{thm:evector}
Let $A\in\mathbb{R}_{\min}^{N\times N}$ be irreducible and define the normalised
matrix $\hat{A}=-\lambda\odot A$.  Consider
$\hat{A}^*$ to be the Kleene star of $\hat{A}$.
\begin{enumerate}
\item If node $i$ belongs to $\mathcal{G}^{cr}(A)$, then
$[\hat{A}^*]_{\cdot i}$ is an eigenvector of $A$.
\item The eigenspace of $A$ is
\begin{equation}
V(A)=\left\{\mathbf{v}\in\mathbb{R}_{\min}^n|\mathbf{v}=\bigominus_{i\in
V^{cr}(A)}a_i\odot[\hat{A}^*]_{\cdot i} \quad\textrm{for
}a_i\in\mathbb{R}_{\min}\right\}.
\end{equation}
\item For $i,j$ belonging to $\mathcal{G}^{cr}(A)$, there exists
$a\in\mathbb{R}$ such that
\begin{equation}
a\otimes[\hat{A}^*]_{\cdot i}=[\hat{A}^*]_{\cdot j}
\end{equation}
if and only if $i$ and $j$ belong to the same MSCS of
$\mathcal{G}^{cr}(A)$.
\end{enumerate}
\begin{proof}
See Theorem 4.5, \cite{heidergott}
\end{proof}
\end{theorem}

So the number of maximal strongly connected components in the critical graph of our min-plus matrix $Q$ determines the number of linearly independent eigenvectors of $Q$.  Consequently we propose that the number of period-1 attractor networks is the same number.  In Figure~\ref{fig:reducednetworks_minplus} there is only one such attractor network because the critical graph of $Q$ consists of one MSCS, namely the circuit between nodes 2 and 3.  In all cases the number of different period-1 attractor networks is bounded by the number of nodes $N$.

%
%We can find proportion of this as follows.  First find the eigenspace -- the number of linearly independent eigenvectors -- and then calculate its proportion when drawn from a uniform distribution (or a distribution of your choice).  E.g. If we let $\mathbf{x}(0)$ be the \textit{integers} in $U[1,10]$, then there are $10^N$ different x(0) possible.  Out of these some will be equivalent, e.g., there are 10 $\mathbf{x}(0)$ that are uniform (all nodes have same value), some will be parallel... The eigenspace includes all vectors that are parallel to the ($\leq N$) eigenvectors.  Worst case: 1 eigenvector (+10 parallel), giving a probability of $10/10^N$ to obtain unique attractor network.  This is a lower bound, since some non-eigenvectors may be iterated to an eigenvector, i.e., contained in the basin of attraction of an eigenvector.  ... Kinda works with $N=3$:  We obtained unique attractor network 53\% of the time.  For $N=3$, above quick calc gives $10/10^3$, which is 10\%...

For the attractor networks with period greater than 1, we will require the following crucial theorem of tropical mathematics, which gives a guarantee of a periodic regime.  First, let $A\in\mathbb{R}_{\min}^{N\times N}$ be irreducible.  The \emph{cyclicity
of $A$}, denoted $\sigma(A)$, is defined as the cyclicity of the
critical graph of $A$.  When the matrix is understood, the cyclicity is also denoted by $\sigma$.
\begin{theorem}\label{thm:Alambdacyc}
Let $A\in\mathbb{R}_{\min}^{N\times N}$ be an irreducible timing dependency matrix with eigenvalue $\lambda$ and cyclicity $\sigma$.  Then there is a $k_\star$ such that
\begin{displaymath}
A^{\odot(k+\sigma)}=\lambda^{\odot\sigma}\odot A^{\odot k}
\end{displaymath}
for all $k\geq k_\star$.
\begin{proof}
See Theorem 3.9, \cite{heidergott}.
\end{proof}
\end{theorem}
Now, post-multiply matrix $A$ from Theorem~\ref{thm:Alambdacyc} by an initial vector $\mathbf{x}(0)$ to obtain 
\begin{equation}
A^{\odot(k+\sigma)}\odot\mathbf{x}(0)=\lambda^{\odot\sigma}\odot A^{\odot k}\odot\mathbf{x}(0)
\end{equation}
Using the original system equation~(\ref{equ:minplusSystem}) along with (\ref{equ:powers_minplusA}), we can simplify this to 
\begin{equation}
\mathbf{x}(k+\sigma) = \lambda^{\odot\sigma}\odot\mathbf{x}(k)
\end{equation}
which is just the statement of a periodic regime.  That is, given an irreducible timing dependency matrix $A$ with cyclicity $\sigma$ and eigenvalue $\lambda$, any periodic regime will have period $\sigma$ and cycletime $\lambda$.   Recall that the period is actually dependent on $\mathbf{x}(0)$, so this period will more accurately be a factor of $\sigma$, but the important point to note is the following theorem.
\begin{theorem}\label{thm:attractornetwork}
Let $\mathcal{G}$ be a strongly connected network having timing dependency matrix $A$ with cyclicity $\sigma$.  Then there is a $k_\star$ such that
$$ \mathcal{G}_r(k+\sigma) = \mathcal{G}_r(k) $$
for all $k\geq k_\star$.
\begin{proof}
This follows from Theorem~\ref{thm:Alambdacyc}, whose conditions are met by $A$.  A corollary of Theorem~\ref{thm:Alambdacyc} is that vectors in a periodic regime separated by $\sigma$ will be parallel because they will differ by the same constant, namely the eigenvalue $\lambda$ of $A$.  Therefore, epochs in the regime separated by the $\sigma$ will have the same reduced network. 
\end{proof}

\end{theorem}
Observe then that the cyclicity of $Q$ for the network of Figure~\ref{fig:regular3nbhd3node} is $\sigma = 2$ (the length of the critical circuit through nodes 2 and 3).  Therefore, the period of any first-input periodic regime will be no larger than 2.  Together with the eigenspace of $Q$ having dimension 1, we can now see why we obtain only two attractor networks in Figure~\ref{fig:reducednetworks_minplus}.  The attractor network (a) is following on from a periodic regime with period 2.   

It would follow that the eigenspace of $Q^{\odot 2}$ would enumerate the different types of attractor networks of period 2.  Without calculation and observing that there is only one type of period-2 attractor network in Figure~\ref{fig:reducednetworks_minplus}, we claim that the eigenspace of $Q^{\odot 2}$ also has dimension 1.

Note that the concept of an eigenspace can be extended for the case where the timing dependency matrix $A$ is not irreducible.  There is a little more work involved, but the periodic regime is guaranteed \cite{fahim2017generalization}.  Thus, Theorem~\ref{thm:attractornetwork} is also true for a network that is not strongly connected.

\section{Advantages of reducing the maxmin-$\omega$ system}\label{sec:advantages}
Reduced networks of the first-input system suggest some advantages for reducing the network for the many-inputs system, i.e., the general maxmin-$\omega$ system $\mathcal{M}$.  One of the main motivations for constructing reduced networks is that these networks can replace the original network.  Practically, $\mathcal{M}$ can then be replaced by an all-inputs system on the attractor network instead.  This is intuitive - only the affecting nodes affect the future state of a node $i$, so it is equivalent to node $i$ accepting all inputs from $\mathcal{A}_i(k)$.  Since it is the dual to the all-inputs system -- bringing with it the readily applicable theory of tropical mathematics that we have highlighted thus far -- the first-input system is already as computationally efficient as the all-input system on the same network.  Consequently, it seems like a redundant exercise to reduce the first-input system, with the benefits being felt more for reducing the many-inputs system (see next).

However, it turns out that reducing the first-input system does actually speed up computations, as we now elaborate.  In a first-input system, we are concerned with finding the elementary circuits with smallest average weight.  The bulk of the work is contained in enumerating such circuits, and the best algorithm we are aware of takes ${O((|V|+|E|)(c+1))}$ operations to do this, where $c$ is the number of elementary circuits \cite{johnson1975finding}.  Of course, reducing the network $\mathcal{G}=(V,E)$ decreases the number $|E|$ of edges and, in a first-input system, all but one incoming edge is removed at each node such that the neighbourhood size $m$ of each node in the reduced network $\mathcal{G}_r$  is 1.   Denote the number of edges in $\mathcal{G}_r$ by $|E|_r$.  Then, $|E|_r = |V|$, the number of nodes.  In fact, some inputs may arrive simultaneously, and so, as in attractor network (a) of Figure~\ref{fig:reducednetworks_minplus}, some edges will not be removed in the process.  Therefore, $m\geq 1$ for some nodes.   So $|V|$ is actually a lower bound for $|E|_r$, a `best case' scenario.

Once found, an all-inputs system can be run on $\mathcal{G}_r$; again, this requires enumerating circuits (and then finding the largest average circuit weight) to find the cycletime, i.e., the asymptotic growth rate of the update times.  Thus, overall, this takes  $O(|V|(c_r+1))$ operations, where $c_r$ is the number of elementary circuits in $\mathcal{G}_r$.  Since it contains fewer edges, $\mathcal{G}_r$ will contain fewer elementary circuits than the original network $\mathcal{G}$, so $c_r<c$.  In summary, the reduced system will be computationally faster to run than the original system.   

The computational complexity increases with $|E|_r$ and therefore $\omega$.  As a result, although any $\omega<1$ will give rise to a reduced system that is faster to run than the original system, the first-input system is the fastest.  In terms of dynamics for applications, the reduced network is the minimal network that produces the same network state behaviour; all additional edges are redundant.  We comment, however, that the practitioner may want to retain these additional edges for stability.  For instance, in a first-input reduced network of one of the applications mentioned earlier, the functional failure of an edge may affect the dynamics significantly since the successor node would then be starved of any input altogether.  This situation improves as $\omega$ increases, though we have just shown that this comes at a higher computational cost.  As $\omega$ tends towards 1, the attractor networks tend towards the original network, and computational cost is closer to the cost of the original maxmin-$\omega$ system.  We propose then, that  $\omega = 0.5$ gives the best system to reduce if one wants an efficient computational procedure whilst retaining some stability in case of edge failure.  This is even more pronounced when we consider that the number of different types of attractor networks is maximized when $\omega = 0.5$.  That is, a large number of different attractor networks (obtained from different initial conditions $\mathbf{x}(0)$) give a large number of options for the reduction of the maxmin-$\omega$ system.  This is especially relevant for applications since it gives some flexibility as to the type of reduced network one can use; we expand on this in the next section.

%Alternatively, we can think of computing the eigenvalue of a max-plus matrix.  Of the commonly-used algorithms for computing the eigenvalue of a max-plus matrix, perhaps the best known is Karp's algorithm [13, 4, 9]. This finds the maximum cycle mean of an $n\times n$ matrix in $O(n|E|)$ operations where $|E|$ is the number of edges of the communication graph associated with the matrix.

\section{Reduced networks: many-inputs}\label{sec:manyinputs}
We now move on to the general case $0<\omega\leq1$ and relax the minimum constraint so that more inputs are required before nodes update their states.  In contrast to a simple contagion, this case might model a complex contagion, where individuals in a social network require social reinforcement to adopt some fad or information supplied by their neighbors.  We use the network of Figure~\ref{fig:regular3nbhd3node} again and, to set the scene, generate asymptotic quantities for all $\omega$.  For such a size 3 network, there can only be three maxmin-$\omega$ systems: first-input, many-inputs (here called second-input) and all-inputs.  For any initial condition $\mathbf{x}(0)$, Section~\ref{subsec:asymptotics} tells us that the cycletime will be unique.  Thus, we obtain the increasing cycletimes 3, 6.5, and 9 for the maxmin-$(1/3)$ (first-input), maxmin-$(2/3)$ (second-input), and maxmin-$(3/3)$ (all-inputs) systems respectively.  The first-input and all-inputs cycletimes were already found, and it fits intuition to see that the second-input cycletime lies between 3 and 9.

\begin{figure}[!hbp]
\centering
  \includegraphics[scale=0.35,trim=0 410 10 100,clip]{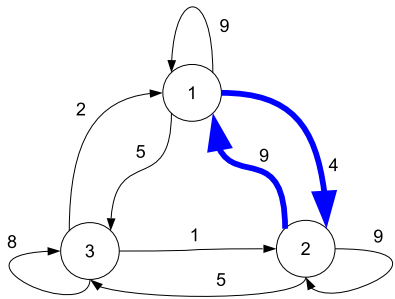}
  
  \caption{Critical circuit highlighted (as thick, blue edges) for a maxmin-$(2/3)$ system on the network of Figure~\ref{fig:regular3nbhd3node}.  This circuit gives cycletime 6.5.  (Color online.)}\label{fig:criticalcircuitsduality}
\end{figure}

Now, we complete the work of Section~\ref{subsec:features} by obtaining statistics of attractor networks for all maxmin-$\omega$ systems on this small network.   For all nodes $i$, the initial state $x_i(0)$ is, again, taken from the uniform distribution between 1 and 10.  Table~\ref{tab:attractornetworks} shows the attractor networks and their proportions for the first-input and second-input cases.  Of course, the all-inputs system results in the same `reduced' attractor network as the original network, so this is not shown.
\begin{table}[!hbp]
\centering
\begin{tabular}{ | c l | c l |}
\hline
\small{first-input }& \small{proportion}& \small{second-input} & \small{proportion} \\
\hline

\includegraphics[scale=0.2,trim= 315 265 400 215 , clip]{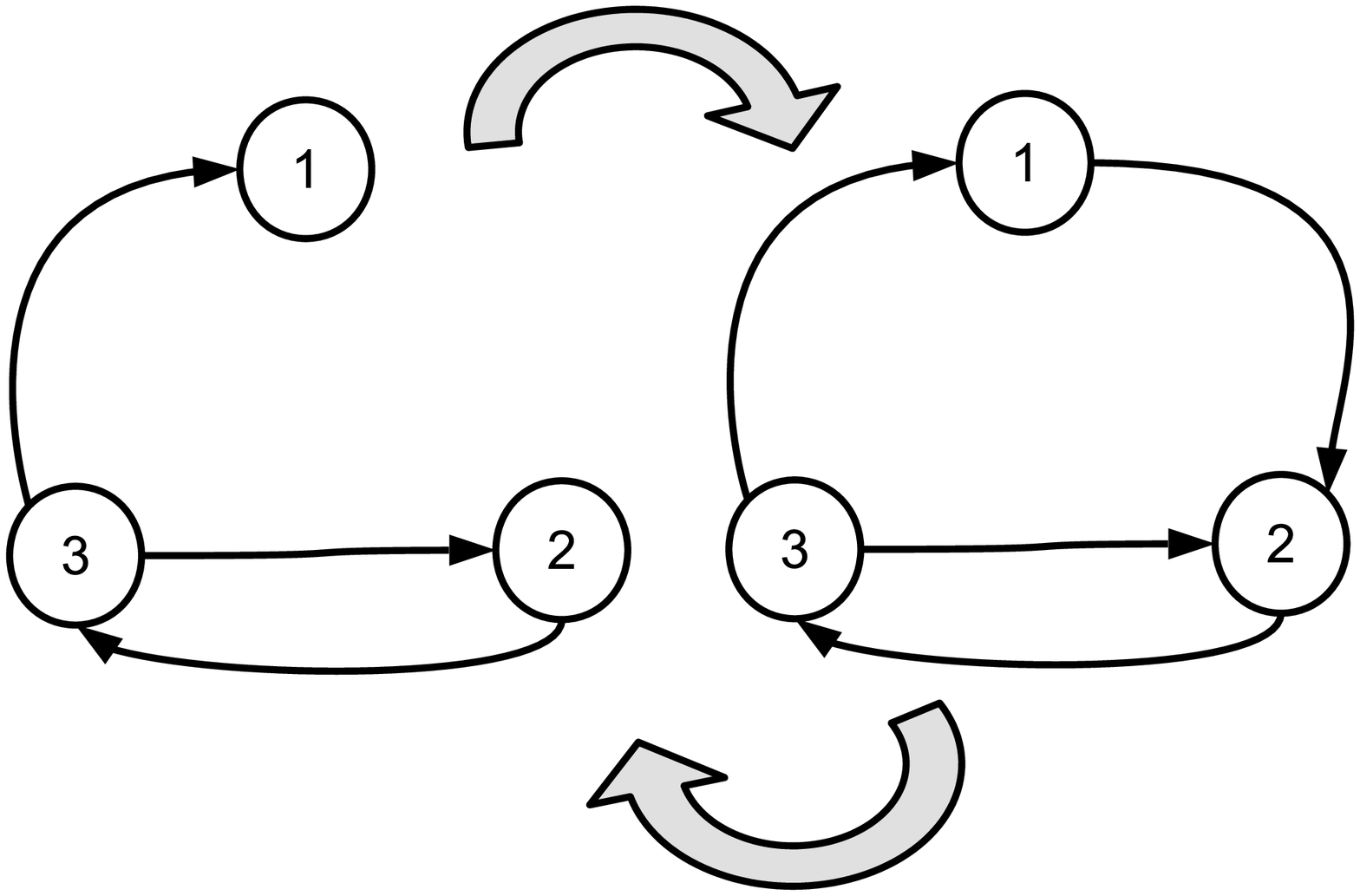} &  47 & \includegraphics[scale=0.2,trim= 315 265 400 215 , clip]{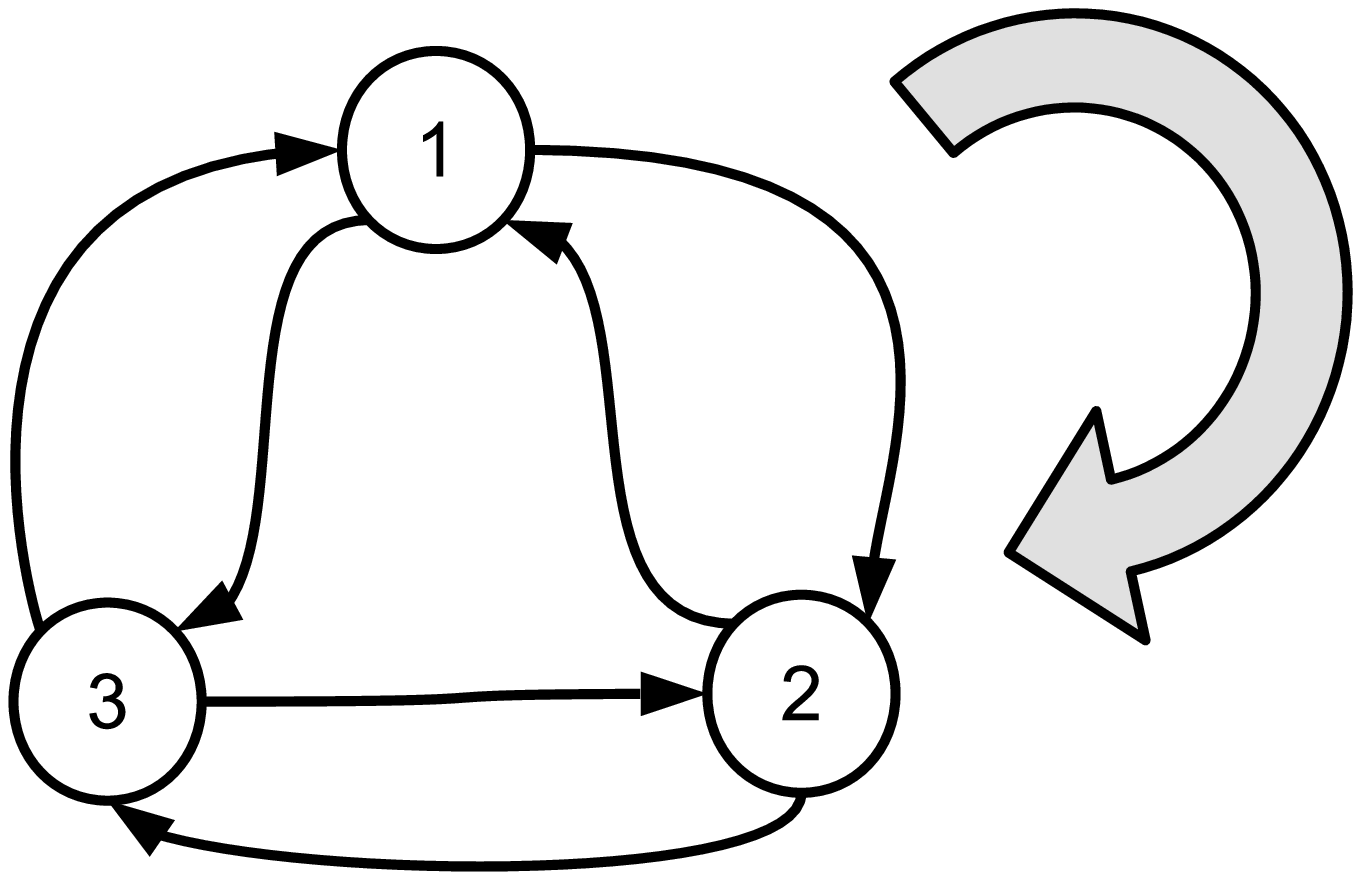} &  34 \\
\hline
\includegraphics[scale=0.2,trim= 315 265 400 215 , clip]{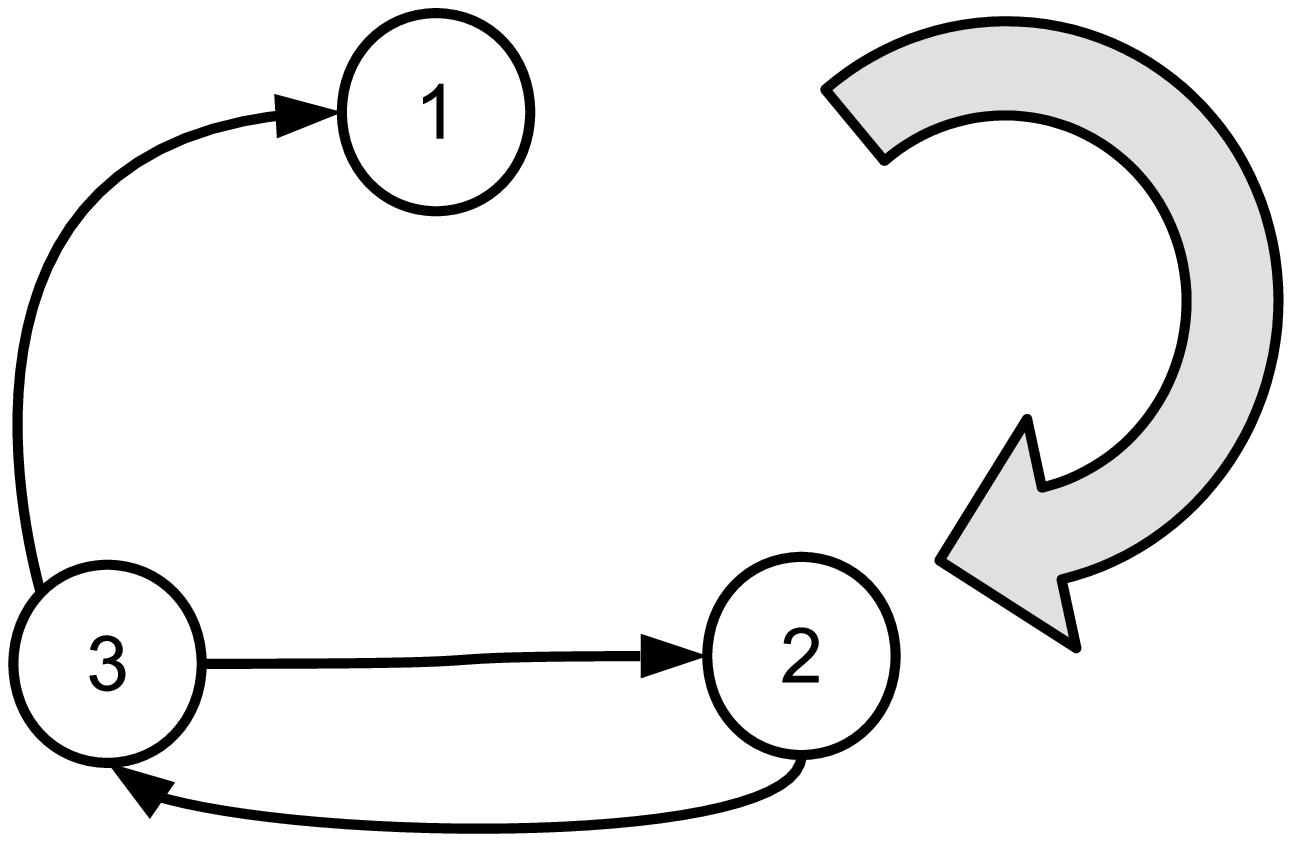} &  53 &  \includegraphics[scale=0.2,trim= 315 265 385 160 , clip]{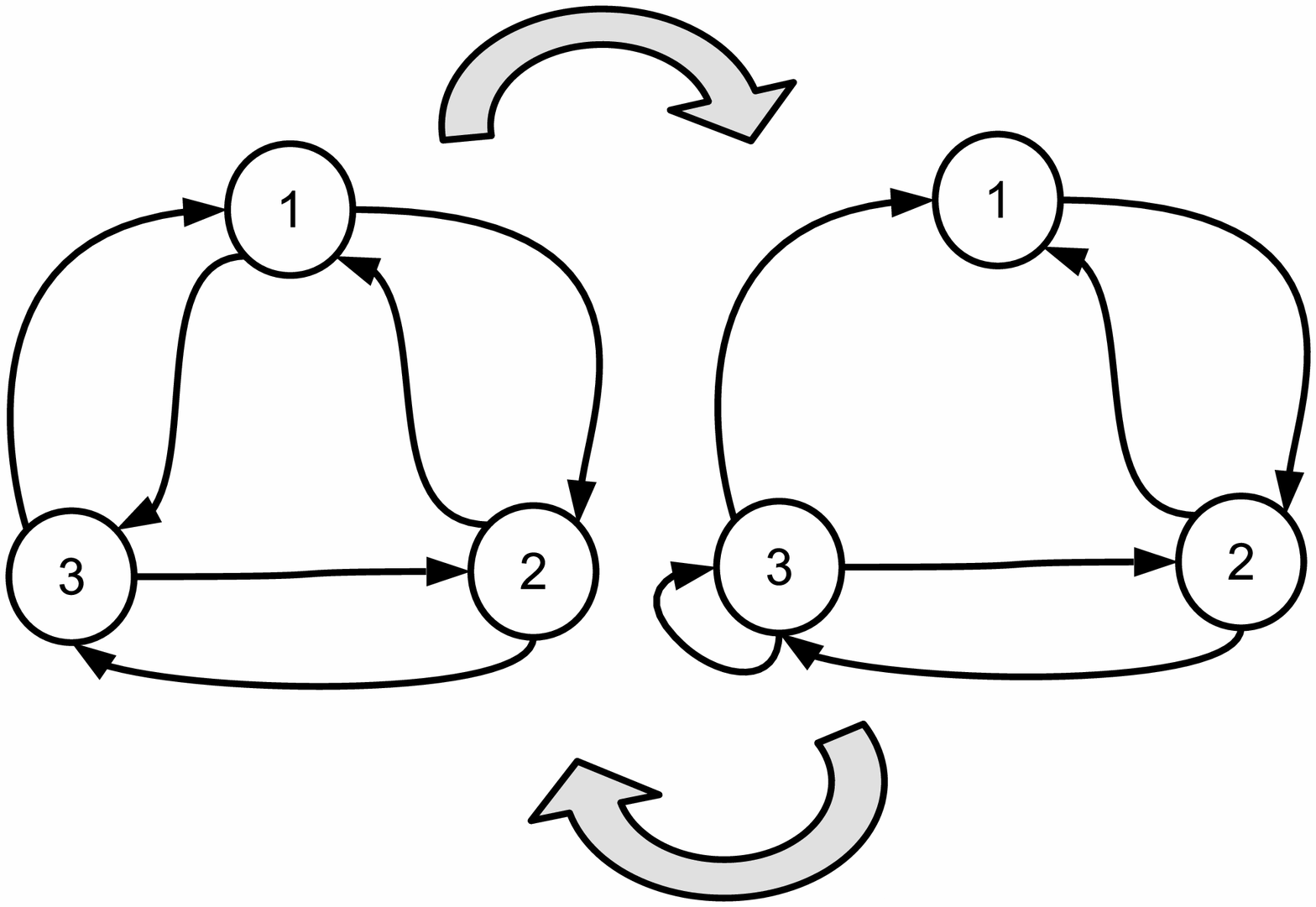} &  20.1\\  \hline
 &   &  \includegraphics[scale=0.2,trim= 300 265 385 190 , clip]{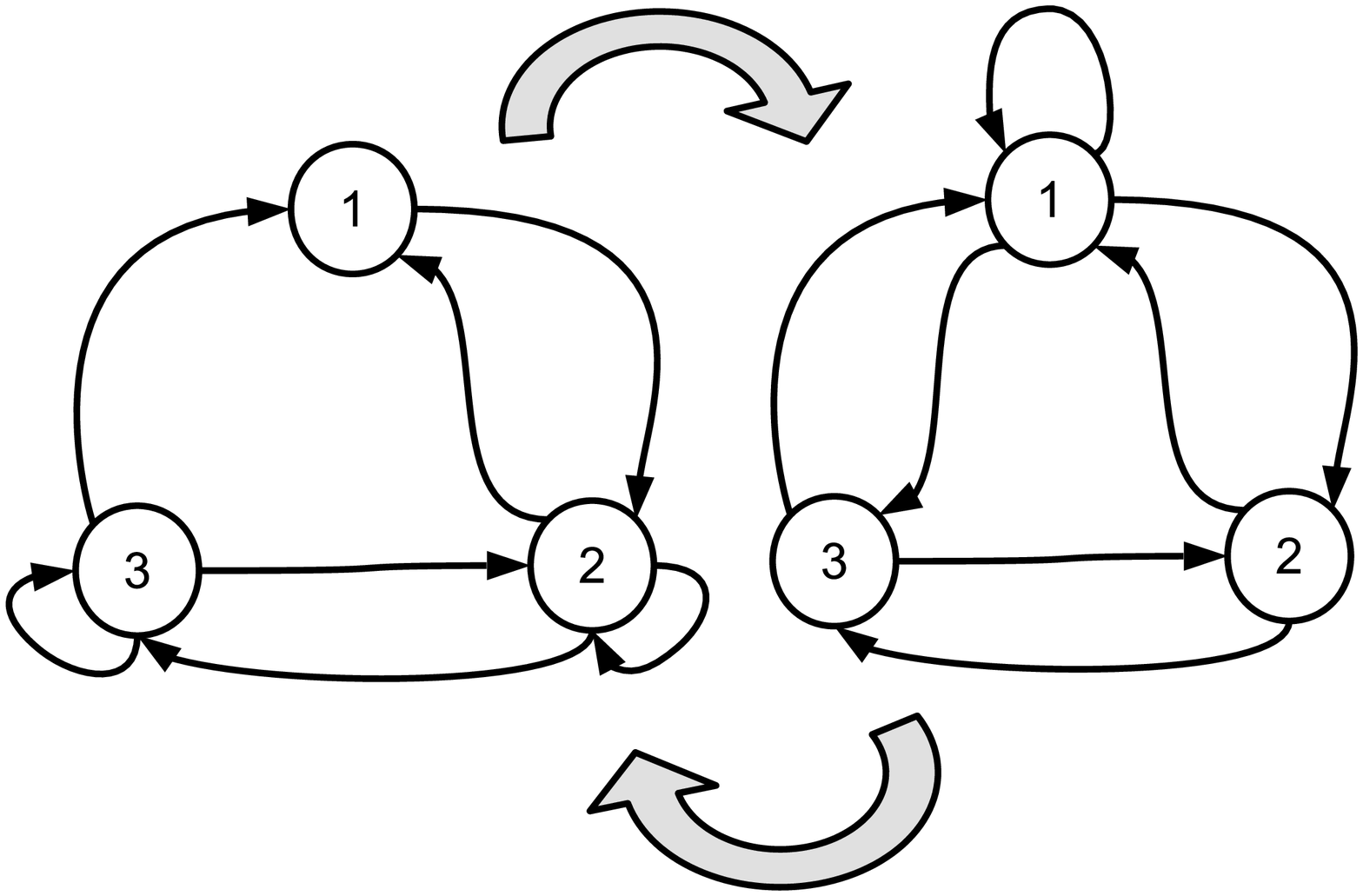} &  40.3\\  \hline
  &   &  \includegraphics[scale=0.2,trim= 300 265 385 190 , clip]{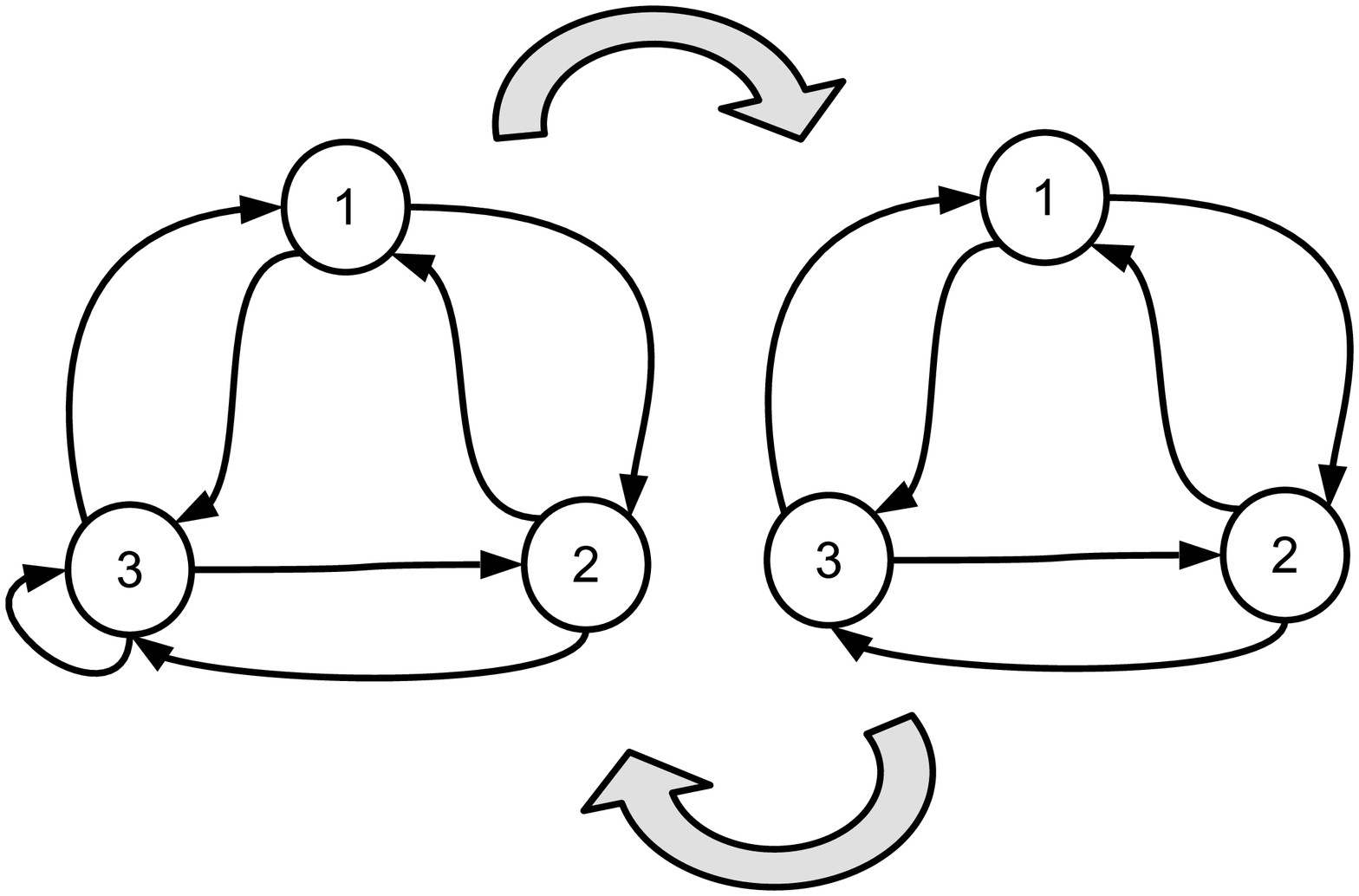} &  5.6\\ \hline

\end{tabular}
\caption{Attractor networks collated from 10000 different initial conditions $\mathbf{x}(0)$ for the maxmin-$\omega$ system on the 3-node network of Figure~\ref{fig:regular3nbhd3node}.  Each element of $\mathbf{x}(0)$ is taken from the uniform distribution between 1 and 10.  The proportion is given as a percentage of the 10000 runs.}
\label{tab:attractornetworks}
\end{table}
Although we found the cycletime for tropical systems as the largest and smallest average circuit weights, it is impossible to do something similar for the second-input system (see \citep{ELPthesis} for algebraic reasons).  Therefore, we enumerate the circuits and find those circuits whose average weight matches the cycletime 6.5.  In this case, this turns out to be the length-two circuit between nodes 1 and 2 (See Figure~\ref{fig:criticalcircuitsduality}).   Notice that we first have to obtain the cycletime here by running the system from any initial condition.  This process must consequently be more computationally intensive than for tropical systems.   We notice, however, that this critical circuit is contained in all four second-input attractor networks of Table~\ref{tab:attractornetworks}.

\subsection{Features of maxmin-$\omega$ attractor networks}
Along with the first-input results of Section~\ref{subsec:features}, Table~\ref{tab:attractornetworks} is indicative of similar studies that we performed on various other small networks.  It allows us to formulate some conjectures on the features of reduced networks.  Thus, consider the general maxmin-$\omega$ system $\mathcal{M}$ on a network $\mathcal{G}$ that produces the periodic orbit $\mathcal{O}$ of reduced networks.  Then
\begin{enumerate}
\item Each network in $\mathcal{O}$ contains critical circuits of $\mathcal{M}$. 
\item If $\mathcal{O}$ has period $g=1$, then the edges of this network satisfy the condition that the neighborhood size of node $i$ must be equal to $\omega|\mathcal{N}_i|$ (for all $i$).  If $g>1$, then at least one node in at least one of the networks in $\mathcal{O}$ will have neighborhood size greater than $\omega|\mathcal{N}_i|$.
\item If $g=1$, then the average weight of a non-critical circuit in the attractor network $\mathcal{G}_r$ is strictly less than the cycletime of nodes in that circuit.  If $g>1$, then at least one of the $g$ attractor networks contains at least one elementary circuit with average weight greater than the cycletime of nodes in that circuit.
\item  If $g=1$, then the cycletime of node $i$ is equal to the largest average circuit weight from all maximal strongly connected subgraphs of $\mathcal{G}_r$ that are predecessors of $i$. 
\end{enumerate}
This points towards the following algorithm for constructing attractor networks of period 1.  Along with the critical circuits, the algorithm adds further edges to ensure that the neighborhood size of each node in $\mathcal{G}_r$ is equal to $\lceil{\omega|\mathcal{N}_i|}\rceil$, whilst checking that the average weight of circuits in $\mathcal{G}_r$ do not contradict the cycletime condition, as required.

\begin{algorithm}[Constructing an attractor network of period 1] \label{alg:attractornetwork}

Given the maxmin-$\omega$ system $\mathcal{M}$ on a network $\mathcal{G}=(V,E)$ of size $N$.  Start with a network $\mathcal{G}_r$ of all nodes in $V$ but no edges.
\begin{enumerate}
\item Choose any $\mathbf{x}(0)$ and run $\mathcal{M}$ on $\mathcal{G}$ to obtain a periodic regime.  Note the cycletime vector $\chi(\mathcal{G})$.
\item To $\mathcal{G}_r$, add the critical graph $\mathcal{G}^{cr} = (V^{cr}, E^{cr})$, i.e., circuits in $\mathcal{G}$ that have the same average weight as the cycletime vector $\chi$. 
\item For $i = 1,\ldots,N$, order the set $E^{i} = \{(j,i)\in E|j=1,\ldots,N\}$ of all incident edges from $E$ on node $i$ by edge weight, smallest first. 
\item Choose a permutation $V_p$ of the nodes $1,\ldots,N$.  (E.g., for $N=4$,  $V_p = \{2,3,1,4\}$.)
\item For nodes $i = V_p$,

While the number of incident edges from $E_r$ on node $i$ is less than $m=\lceil{\omega|\mathcal{N}_i|}\rceil$, 
\begin{itemize}
\item For edges $a=1,\ldots,|E^{i} |$,
\begin{itemize}
\item Add edge $a\in E^{i}$ to $E_r$.
\item If there are circuits in $\mathcal{G}_r$ whose average weight is greater than the cycletime of nodes in that circuit, then remove edge $a$ from $E_r$.  
\begin{itemize}
\item If $a = |E^{i} |$ (i.e., we have exhausted all edges in $E^{i}$ and number of incident edges is still less than $m$), then exit and move to Step 4 to choose a different ordering of nodes and run Step 5 again.
\end{itemize}
\end{itemize}
\end{itemize}

\end{enumerate}
\end{algorithm}
Note that, for a first-input system, Step 1 can be omitted, since the cycletime can be deduced as the smallest average circuit weight in $\mathcal{G}$.   In comparison to employing our statistical study -- where we generate many thousands of initial conditions $\mathbf{x}(0)$ and collate the different attractor networks -- the algorithm is computationally more complex for a many-inputs system, especially when it only gives a period-1 attractor network.

%Perhaps faster to consider a particular permutation $V_p$ of nodes to speed up convergence.  E.g., $V_p$ could ordered according to node in-degree.  (Otherwise, there are $N!$ possible permutations!)

The concept of an eigenspace is readily understood for a tropical maxmin-$\omega$ system $\mathcal{M}$ since it is represented by the timing dependency matrix $P$ (or $Q$).  Let us loosely generalise this concept to say that an eigenvector $\mathbf{v}$ and eigenvalue $\lambda$ of $\mathcal{M}$ satisfy 
\begin{equation}
\mathcal{M}(\mathbf{v}) = \lambda\otimes\mathbf{v} \quad .
\end{equation}
Then, we hypothesize that the eigenspace of $\mathcal{M}$ determines the number of different period-1 attractor networks.  Similarly, the eigenspace of $\mathcal{M}^{g}$ determines the number of different period-$g$ attractor networks.  

For period $g>1$, we must consider the fact that some circuits in at least one of the attractor networks of the periodic orbit will contain a circuit with average weight larger than the cycletime.  This means that, if it was considered on its own as an attractor network of period 1, its cycletime would not match the cycletime as measured by max-plus algebra on this reduced network.  

Notice that some attractor networks will contain nodes whose neighborhood size is larger than $m$.  This is the result of inputs arriving at the same time -- something we call ``simultaneity" -- and expanded upon in \cite{ELPthesis}.  This means that, even though the eigenspace of the first-input system will give the number of different period-1 attractors, if an attractor network exhibits simultaneity, then it will not be picked up by Algorithm~\ref{alg:attractornetwork}.  It also means we can't easily formulate an algorithm for attractor networks $\mathcal{O}$ of period larger than 1.

Nevertheless, in all cases of $g\geq 1$,  the following must be true. Consider the max-plus adjacency matrices $P^{(1)}_r,P^{(2)}_r,\ldots,P^{(g)}_r$ of each reduced network in $\mathcal{O}$. Without loss of generality, let $P^{(k)}_r$ be the timing dependency matrix of the reduced network on epoch $k$.  Then, in the periodic regime, we have
\begin{equation}
\mathbf{x}(k+1) = P^{(k)}_r\otimes \mathbf{x}(k)
\end{equation}
and, given a state $\mathbf{x}(0)$ in this regime, we can apply this equation recursively to yield 
\begin{equation}
\mathbf{x}(g) = R_r\otimes \mathbf{x}(0)
\end{equation}
where $R_r = P^{(g)}_r \otimes P^{(g-1)}_r \otimes \cdots \otimes P^{(1)}_r$.  Thus, we can calculate update times $\mathbf{x}(k)$ on every $g$ epochs in the periodic regime by applying one max-plus matrix $R\in\mathbb{R}^{N\times N}_{\max}$.   The cycletime of $R_r$ will be $g\chi$, where $\chi$ is the cycletime of the original maxmin-$\omega$ system.  Moreover, the graph $\mathcal{G}(R_r)$ will satisfy the rule that all circuits have average weight less than or equal to $\chi$, as required.

\subsection{Statistics of attractor networks}
The difficulty in constructing attractor networks for period $g>1$ leads us to think again about simply collating attractor networks after running the system $\mathcal{M}$ from many randomly chosen initial conditions $\mathbf{x}(0)$.   The hope here is that taking some large number of different initial conditions will capture all the different attractor networks. 

To generate the attractor networks of Table~\ref{tab:attractornetworks}, we took 10000 different initial conditions; together with the fact that the network size $N$ is small, we feel this is sufficient to capture the full spectrum of attractor networks.  As $N$ increases, we may need to take further initial conditions. 

We can get an idea about what to expect using combinatorics.  For simplicity, consider a network size $N$ where all nodes have the same neighborhood size $n$.  Now consider a many-inputs system, which means that nodes in attractor networks will have neighborhood size $m=\lceil{\omega n}\rceil \geq 1$ (ignoring simultaneity to simplify the thought process).  Each row in the adjacency matrix $P^{(k)}_r$ of an attractor network will therefore have $m$ non-zero elements.  Then there are $^nC_m$ ways to choose these $m$ elements on each row.  As there are $N$ rows, we have that $P^{(k)}_r$ can be in one of $\overline{\mathcal{R}} = \left(^nC_m\right)^N$ forms, i.e., there are potentially $\overline{\mathcal{R}}$ different attractor networks.  Table~\ref{tab:numberofR}  gives the value of $\overline{\mathcal{R}}$ for $N=5$ and its different neighborhood sizes.
\begin{table}[!hbp]
\centering
\begin{tabular}{|rr|ccccc|}
\hline 
     &  &  $m$ & & & &  \\
     &   &   1 & 2 & 3 & 4 & 5 \\
\hline 
$n$ & 5    &  3125 &     100000   &   100000    &    3125     &      1 \\
       &   4   &  1024   &     7776    &    1024     &      1 & \\
       &   3   &  243 &  243 &  1 & & \\
      &    2   & 32 & 1 & & & \\
      &   1  & 1 & & & & \\
\hline

\end{tabular}

\caption{The number $\overline{\mathcal{R}} = \left(^nC_m\right)^5$ of different attractor networks possible on a size 5 network.  The original network neighborhood size is $n$ and attractor network neighborhood size is $m$.}
\label{tab:numberofR}
\end{table}

\begin{figure}[!hbtp]
	\centering
	
	\includegraphics[scale=0.27,trim= 0 0 0 0 , clip]{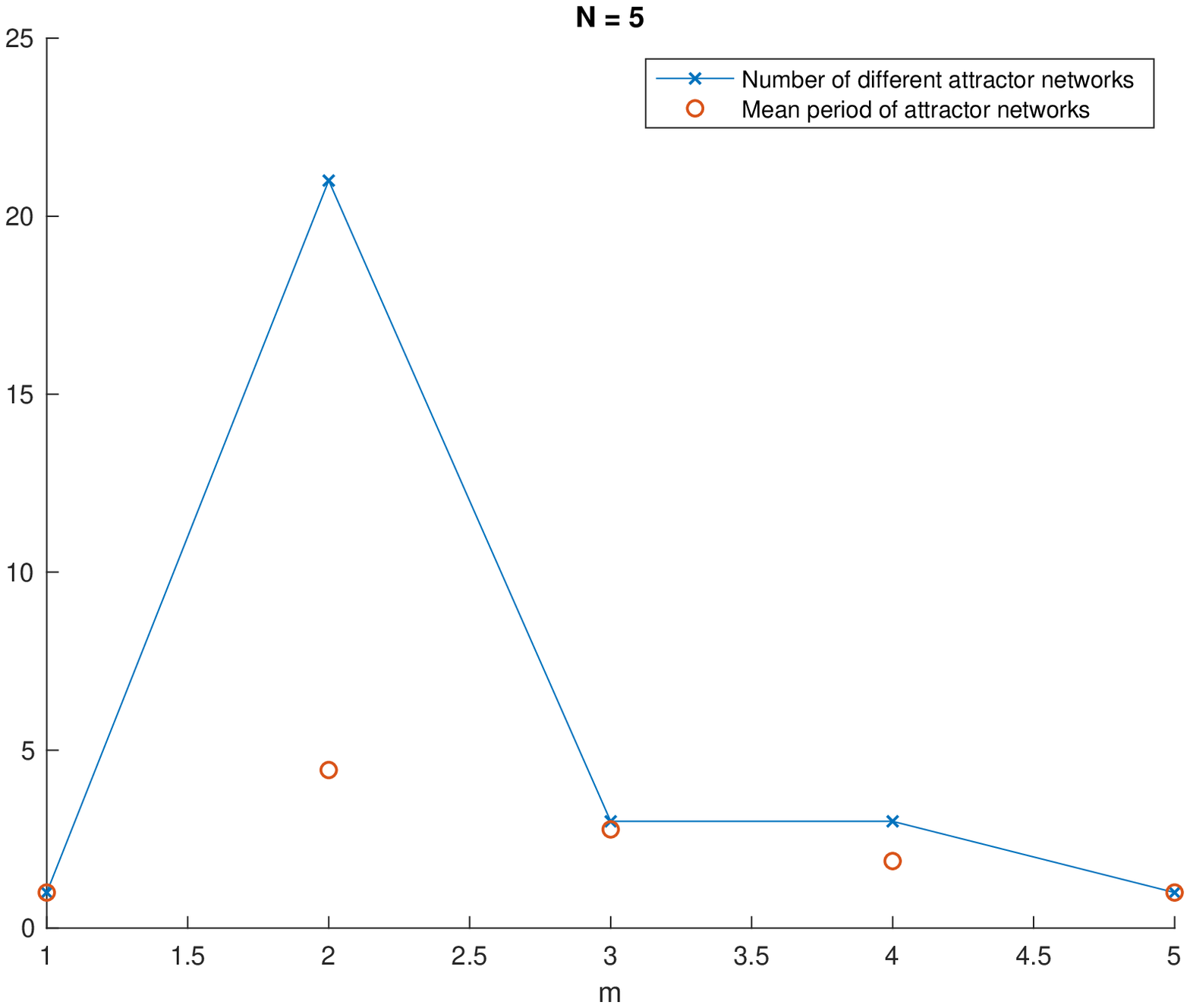} 
	\includegraphics[scale=0.27,trim= 0 0 0 0 , clip]{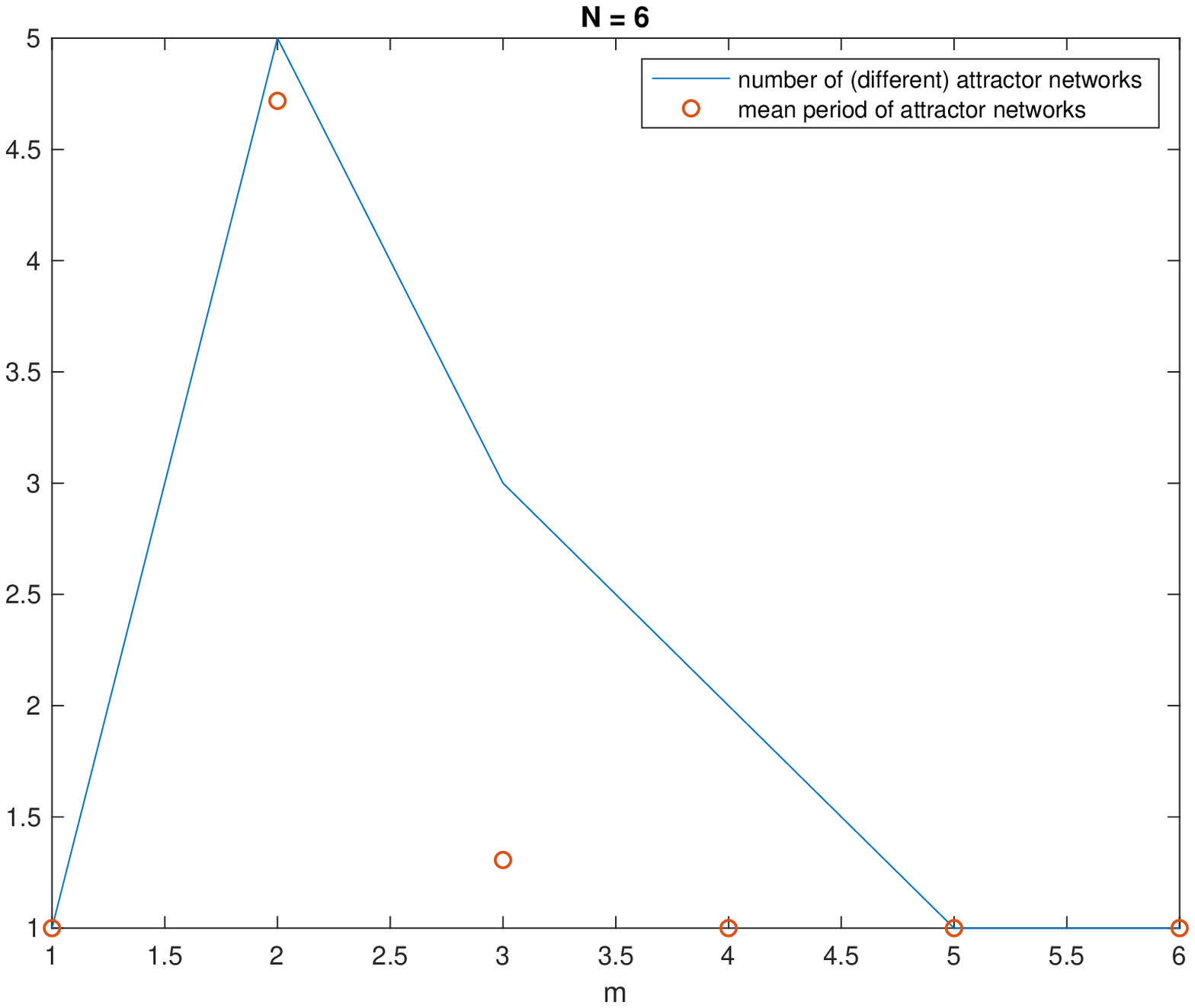}
	\includegraphics[scale=0.27,trim= 0 0 0 0 , clip]{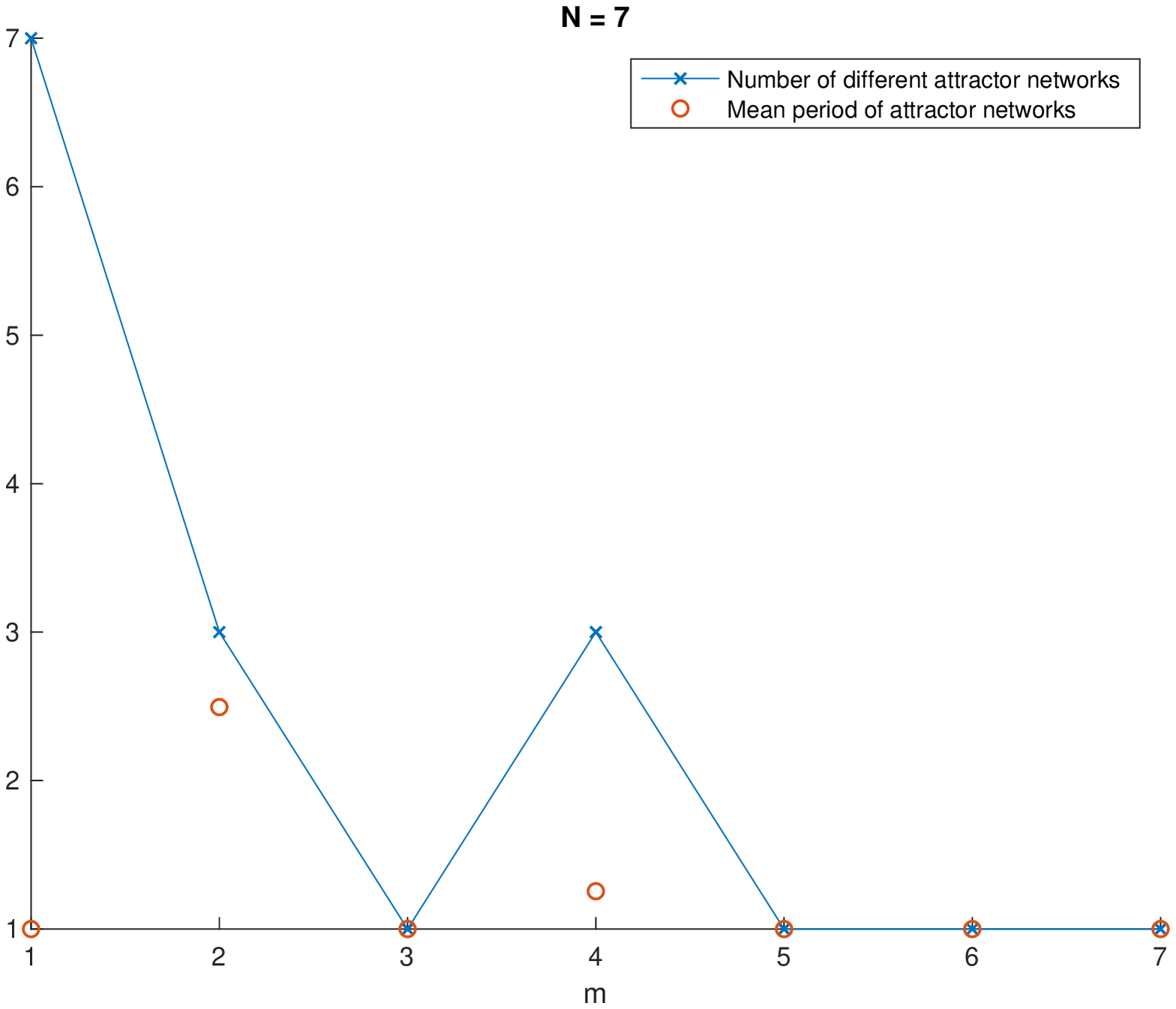}
	\includegraphics[scale=0.27,trim= 0 0 0 0 , clip]{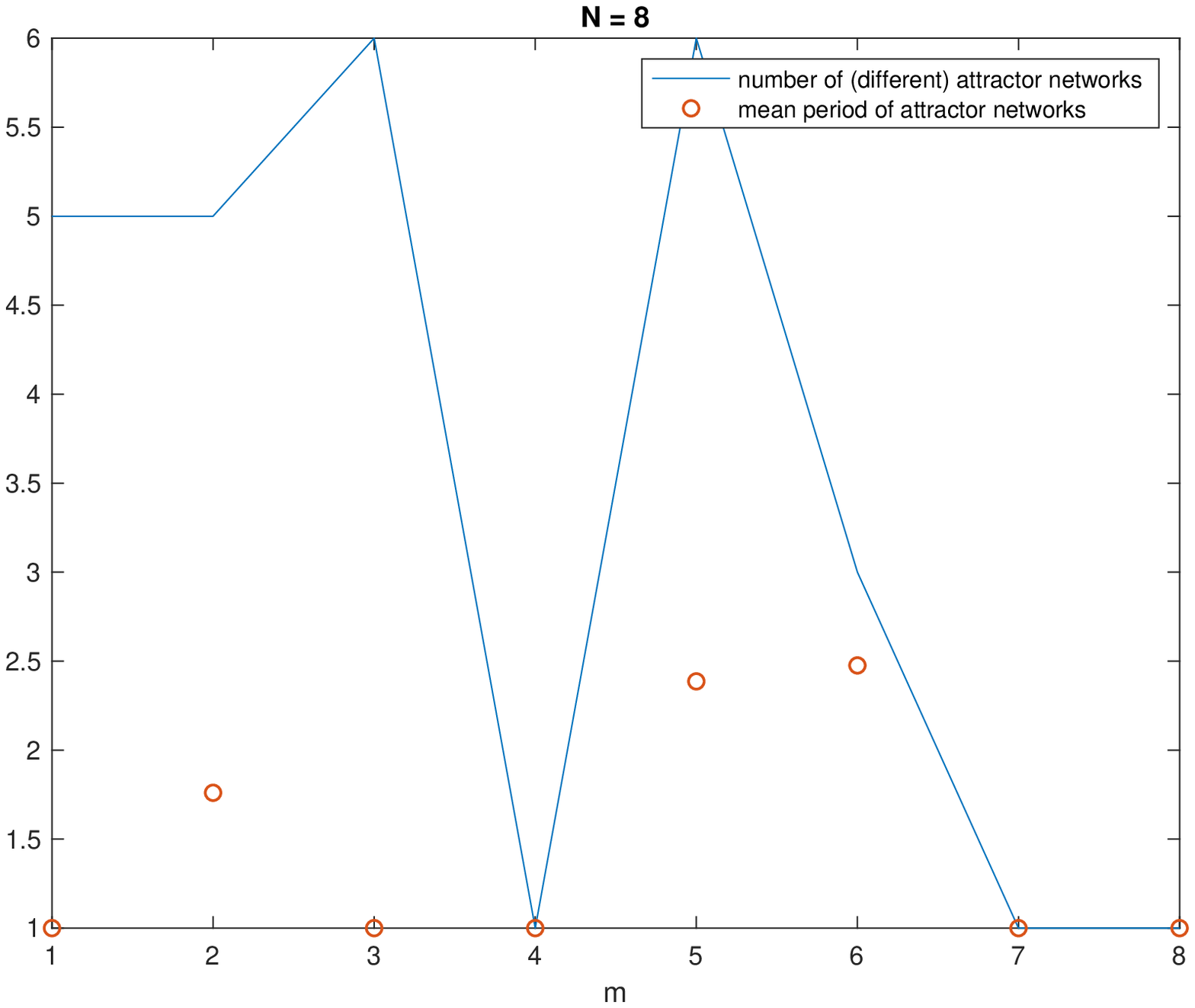}
	\includegraphics[scale=0.27,trim= 0 0 0 0 , clip]{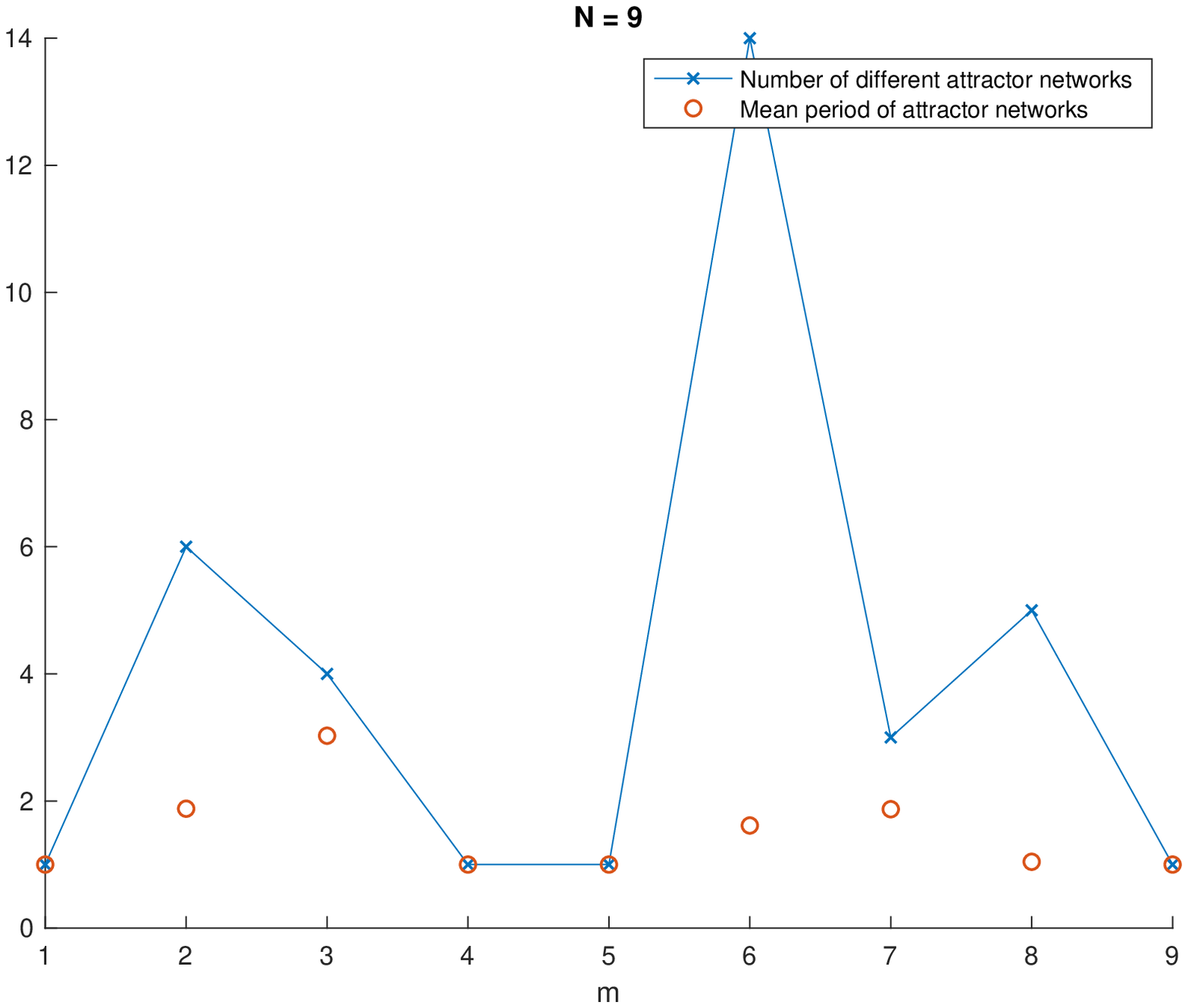}
	\includegraphics[scale=0.27,trim= 0 0 0 0 , clip]{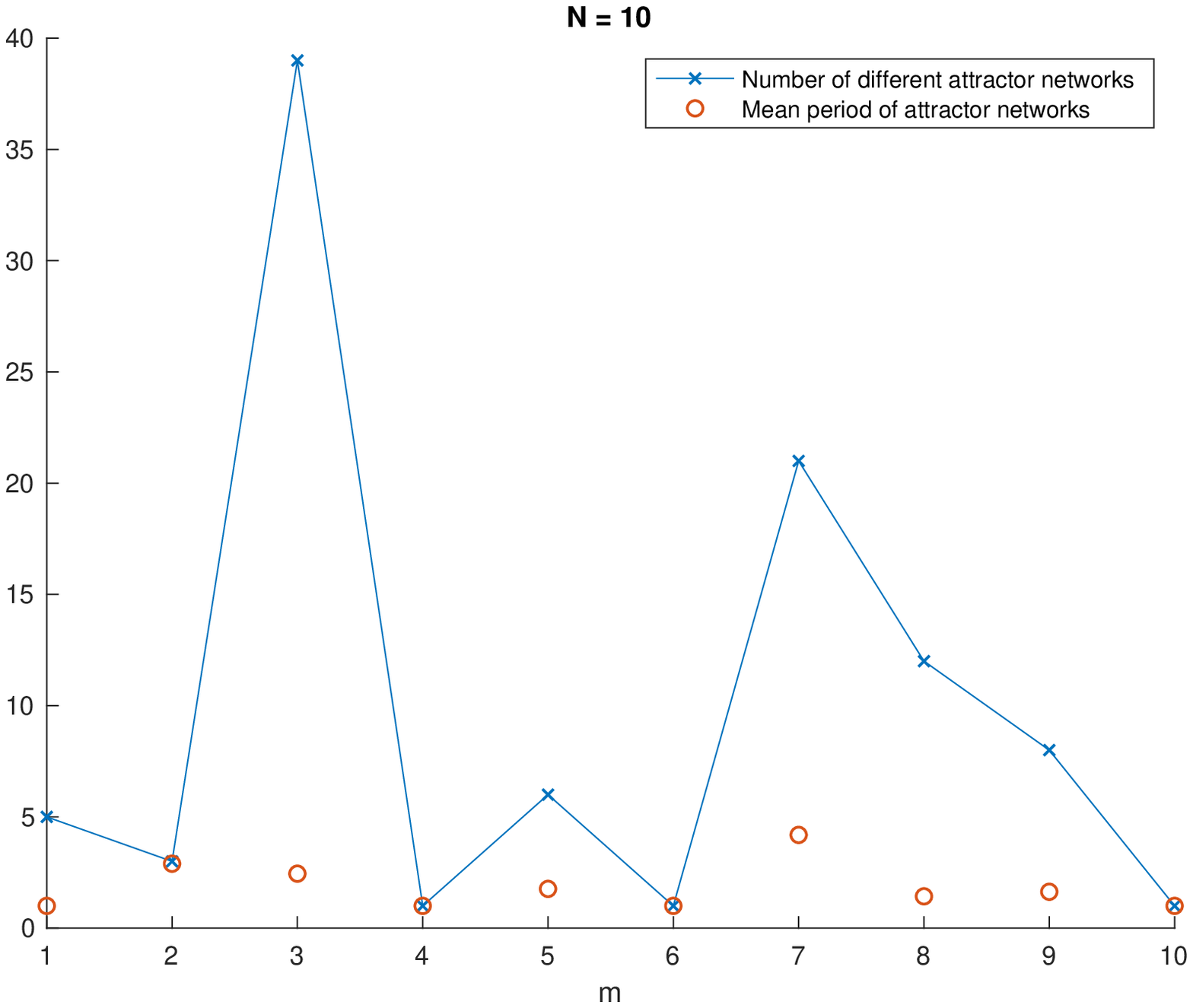}
	
	\caption{The number and mean period of different attractor networks for networks of sizes 5, 6, 7, 8, 9, and 10.  The $x$-axis denotes the maxmin-$\omega$ system by $m$, where $m=\lceil{\omega|\mathcal{N}_i|}\rceil$ for each node $i$.  Edge weights are generated uniformly at random from the integers 0 to 10, where 0 is subsequently interpreted as a lack of an edge.  The numbers are obtained from 10000 initial conditions $\mathbf{x}(0)$ chosen from the uniform distribution between 1 and 10.}
	\label{fig:examplenetworksR}
\end{figure}

\begin{figure}[!hbtp]
	\centering
	
	\includegraphics[scale=0.4,trim= 0 0 0 0 , clip]{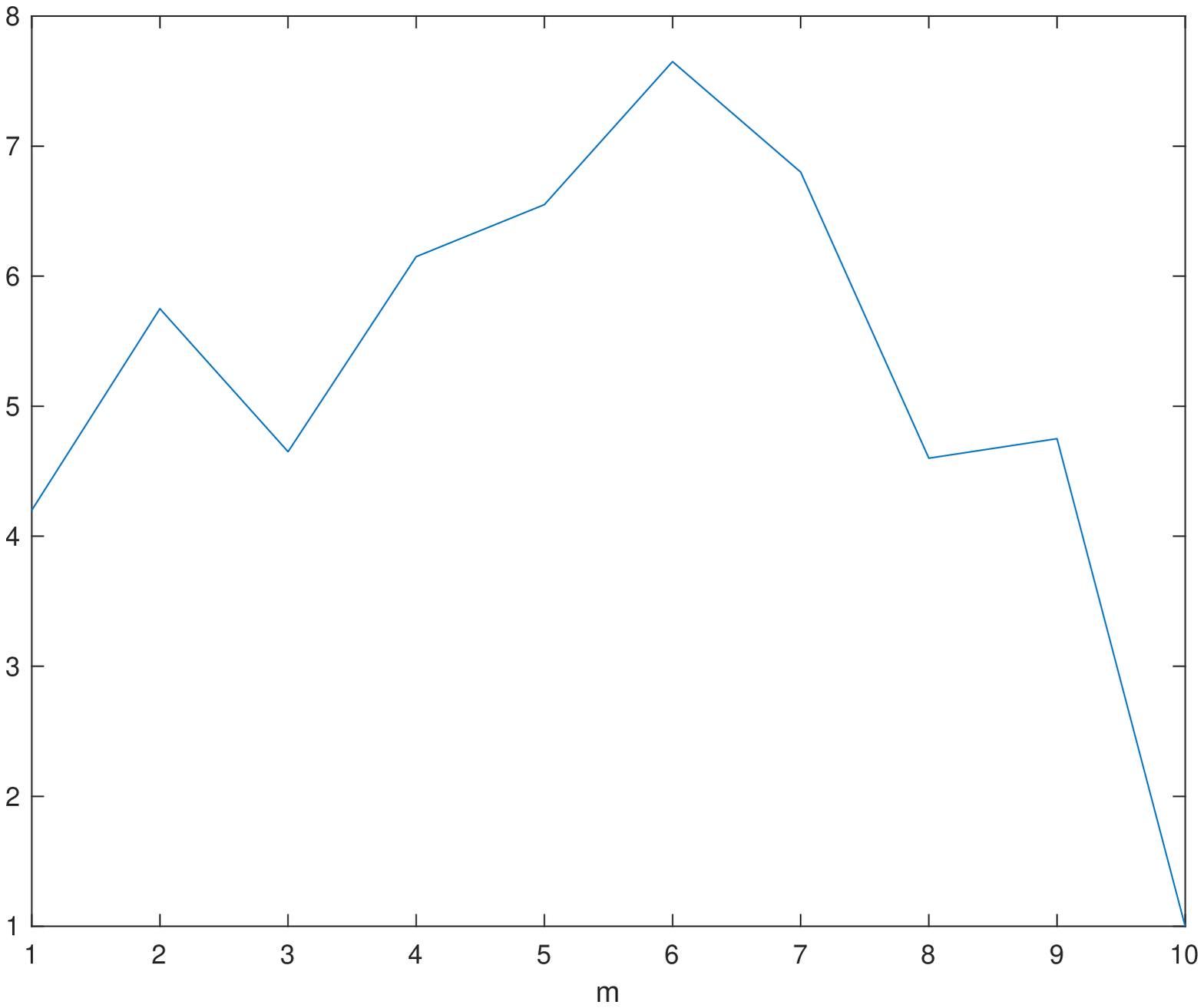} 

	\caption{The mean number of different attractor networks taken from 20 complete networks of size 10, where edge weights in each network are generated from the uniform distribution between 1 and 10.  For each network, the number of different attractor networks are first obtained from 10000 initial conditions $\mathbf{x}(0)$ chosen from the uniform distribution between 1 and 10.  The $x$-axis denotes the maxmin-$\omega$ system by $m$, where $m=\lceil{10\omega}\rceil$ for each node.}
	\label{fig:size10Completenetwork_reducedStats}
\end{figure}

The values of $\overline{\mathcal{R}}$ in Table~\ref{tab:numberofR} may be regarded as a `null model', arising from a complete network where all nodes are connected to each other (even to themselves).  Generating the number of different attractor networks $\mathcal{R}$  for a real network and viewing a deviation from the trend of $\overline{\mathcal{R}}$ with $m$ may reveal something about the network connectivity.  Thus, Figure~\ref{fig:examplenetworksR} shows the number of different attractor networks for six example networks, each of sizes 5, 6, 7, 8, 9, and 10; in term of connectivity, the networks are heavily connected with only a few missing edges.
If the network followed $\overline{\mathcal{R}}$. then we would expect to see a peak in $\mathcal{R}$ at $\omega = 0.5$.  This doesn't occur for most of the networks in Figure~\ref{fig:examplenetworksR}, suggesting a significant deviation from the complete network.  However, we do see this peak in $\mathcal{R}$ when we perform the same analysis on twenty complete networks of size 10 (See Figure~\ref{fig:size10Completenetwork_reducedStats}).

\section{Application to the C. elegans neuron network}\label{sec:apps}
To finish this work, we now present an application of the approaches in this paper.  We have already mentioned some benefits in Section~\ref{sec:advantages}.

\subsection{C. elegans}
Consider the neuron network of the nematode worm, Caenorhabditis elegans (C. elegans).  This network has been studied for almost twenty years now, largely because of its complete mapping and its manageable size; there are 306 neurons, which we interpret as nodes, and neurons send signals to other neurons via synapses.  Thus, we consider the edge weight $\tau_{ij}$ to signify the number of synapses between nodes $j$ and $i$; the network is directed, and asymmetric, so $\tau_{ij}$ is not necessarily the same as $\tau_{ji}$.  The data is taken from \cite{watts1998collective}, ultimately obtained from the seminal work of White et al. \cite{white1986structure}.  We remove isolated nodes, i.e., those nodes with no incident edge; we therefore work with 297 nodes.

\begin{figure}[!hbp]
	\centering
	\includegraphics[scale=0.5,trim= 70 520 50 180, clip]{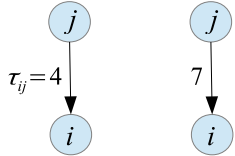} 

\caption{An example of two forms for edge weights.  Left: $\tau_{ij}=4$, which we call the default form (for the neuron network, $\tau_{ij}$ is the number of synapses from neuron $j$ to $i$); right: $\overline{\tau}_{ij} = 7$, which is obtained from (\ref{equ:edgeweights}), and taking $\max_{i,j}{\tau_{ij}}=10$.}
	\label{fig:edgeweights}
\end{figure}

Interpreting edge weights as transmission times suggests two forms: the first form leaves the transmission times unchanged from the source data, and we call it the ``default form", whilst the second form generates new transmission times 
\begin{equation}\label{equ:edgeweights}
\overline{\tau}_{ij} = 1-\tau_{ij}+\max_{i,j}{\tau_{ij}} \ .
\end{equation}
The first form considers that the more synapses there are between nodes, the longer it takes for signals to be transmitted, whilst the second form is where we `reverse' this idea to say that more synapses implies a smaller transmission time (see Figure~\ref{fig:edgeweights}).  

\begin{figure}[!hbp]
	\centering
	
	\includegraphics[scale=0.5,trim= 95 0 70 0, clip]{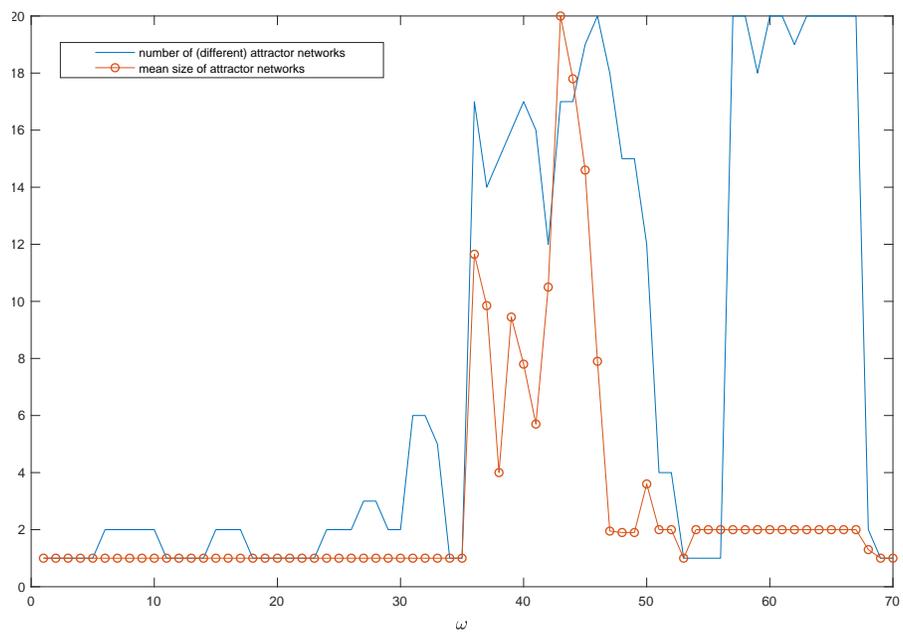} 
\caption{The number and mean period of different attractor networks for the C. elegans neuron network.  The $x$-axis denotes $\omega\in (0,1]$, split evenly into 70 data points.  The numbers are obtained from 100 initial conditions $\mathbf{x}(0)$ chosen from the uniform distribution between 1 and 10.}
	\label{fig:celegansattractorStats}
\end{figure}

Figure~\ref{fig:celegansattractorStats} plots the number $\mathcal{R}$ of different attractor networks for this C. elegans network with default transmission times, together with the periods of these attractor networks.  The main takeaway is that these values are almost constant until a sudden burst at $\omega = 0.5$, from when a variety of attractor networks can be detected.

\begin{figure}[!hbp]
\centering

(a)

\includegraphics[scale= 0.3,trim=115 0 70 10, clip]{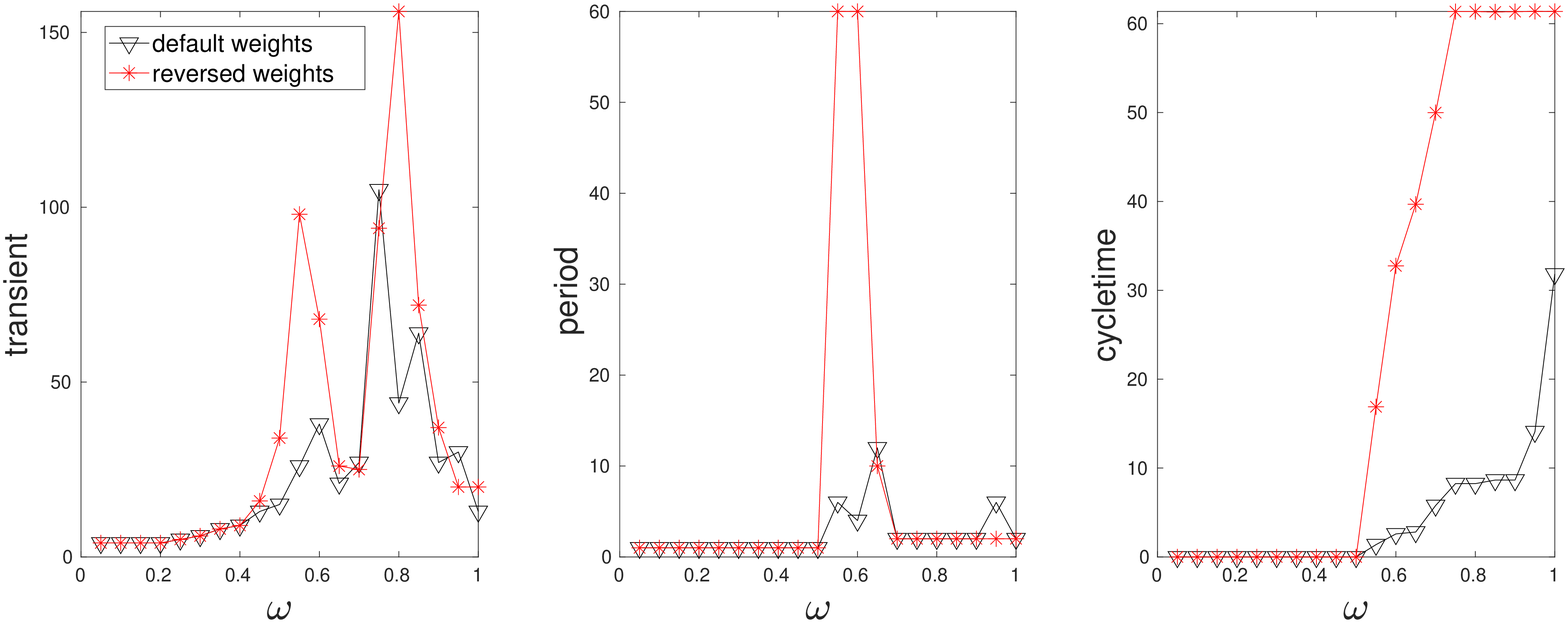}  
	
%(b)	Webster friendship network

%\includegraphics[scale= 0.3,trim=110 0 70 10, clip]{webster}  

(b)

	\includegraphics[scale = 0.33 ,trim=0 0 45 5, clip]{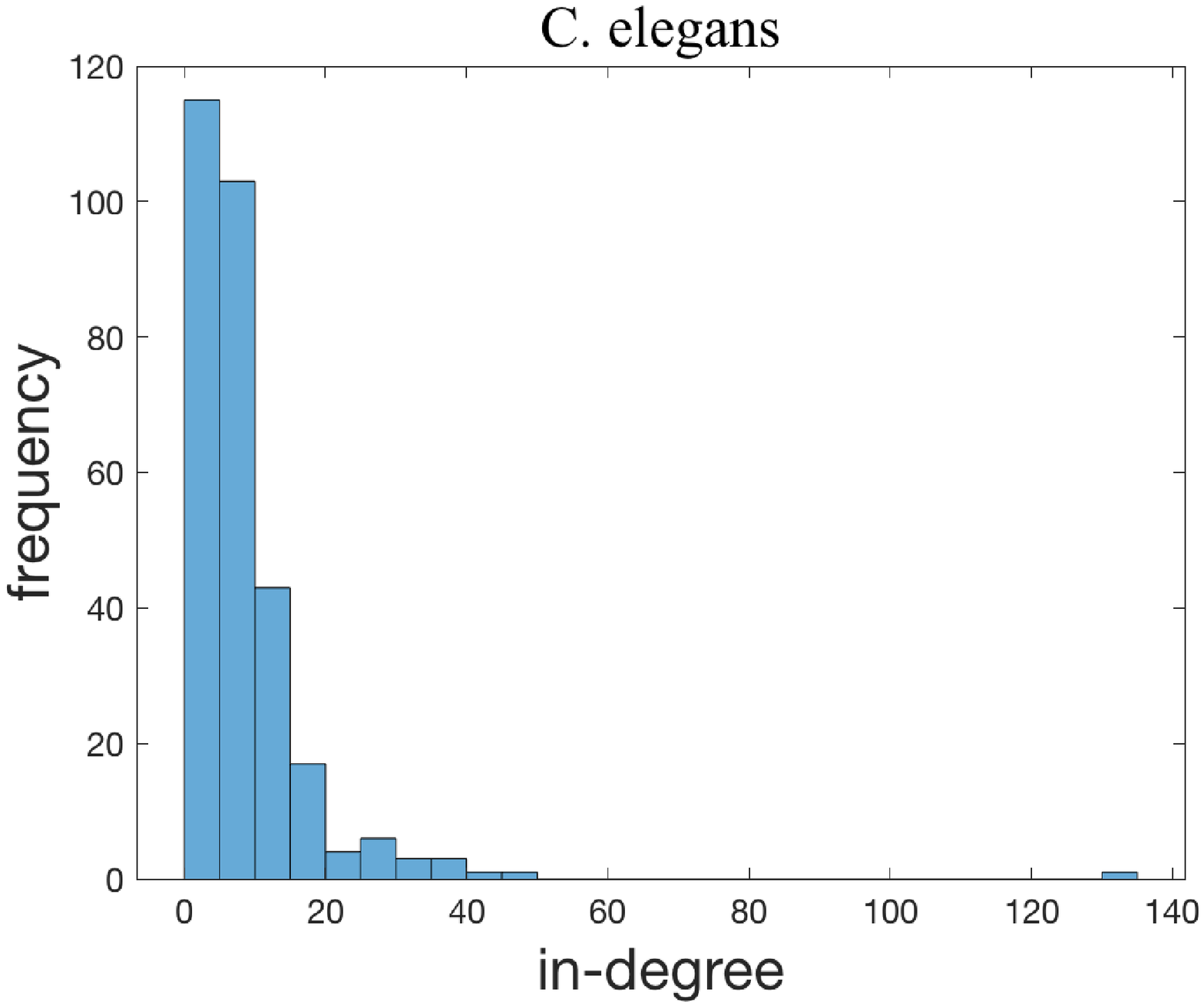} 

\caption{(a) Transient time, period, and cycletime of the maxmin-$\omega$ system as a function of $\omega$ with an underlying network equivalent to the C. elegans neuron network.  Two cases are considered: (i) transmission times are equal to the original network edge weights (``default"), and (ii)  transmission times are the ``reverse" of the original edge weights.  For all cases, we take processing time $\xi_i=1$ for all nodes and initial times $\mathbf{x}(0)=0$.  (b) A histogram of in-degrees for the C. elegans network.  (Color online.)}
\label{fig:realworldresults}
\end{figure}

There is a relationship here with the asymptotics.  We can view the state $\mathbf{x}(k)$ as the firing times of all neurons.  Thus, Figure~\ref{fig:realworldresults}(a) shows the transient time (taken as the largest of all nodal transient times, period (the LCM of all nodal periods), and cycletime (the mean of all nodal cycletimes) of a maxmin-$\omega$ system applied to both cases of this C. elegans neuron network.  The transient time can be seen as the time for the network to settle onto a rhythmic firing pattern, the periodic regime.  The in-degree of a node is equal to the neighborhood size, so that the histogram of in-degrees in Figure~\ref{fig:realworldresults}(b) gives an indication of the network structure: most nodes have very small neighborhoods (size $n= 1$ or $2$) such that, since $m=\lceil \omega n \rceil$, the maxmin-$\omega$ system acts as an all-inputs system ($m=1$) for most nodes.  

Again, we see that, when $\omega$ is approximately 0.5, all three quantities make sudden jumps from almost constant values.  This indicates a percolation-type behavior, that is, the maxmin-$\omega$ system truly takes effect above some critical value $\omega^*$; when $\omega<\omega^*$, all nodes either behave under solely all-inputs or solely first-input dynamics.  For first-input dynamics, $m=1$ so that, solving $m= \omega n $ for $n$ gives $n = 1/\omega$.  For our experiments, we took $\omega$ values starting at 0.05, incremented by 0.05 up to 1; when $\omega=0.05$, $n =20$, whilst $n=2$ for $\omega = 0.5$.  Therefore, for $\omega\leq0.5$, all nodes with neighborhood size 2 or smaller (which is most nodes) behave under min-plus dynamics.  For even smaller values of $\omega$, nodes with larger neighborhood sizes (up to 20) are governed by first-input.  Note also that first-input is equal to all-inputs when $n =20$.

These results, together with the arguments put forth in Section~\ref{sec:advantages}, suggest that heavy-tailed networks of the type of C. elegans may operate more robustly under maxmin-$\omega$ dynamics when $\omega$ is approximately 0.5.

\section{Conclusion}\label{sec:conc}
We have presented maxmin-$\omega$, a new model for asynchronous dynamics on networks.  The asymptotic behavior of this model can be represented by tropical mathematics, which, in turn, suggests a novel approach to model such applications as epidemic spreading and neuron network firing.

For a given network, the attractor networks of a maxmin-$\omega$ system $\mathcal{M}$ can be found by simply running $\mathcal{M}$ from many initial conditions.  Collating the different types of attractor networks shows that they must all contain the critical circuits of $\mathcal{M}$.  Consequently, we have presented an algorithm to find an attractor network if it has period-1; this requires no iterations of $\mathcal{M}$.  The downside is that, if $\mathcal{M}$ is a many-input system, we would need to run the system from some initial condition $\mathbf{x}(0)$ to retrieve its cycletime and then match this cycletime to the average weight of circuits in the network; these circuits are then the critical circuits of the maxmin-$\omega$ system.  It's worth noting that even this slower method is better than a well-known method that uses the duality of max- and min-plus algebra; this method -- better known as the Duality Theorem -- attempts to find the cycletime of $\mathcal{M}$ by first representing it as a max-min-plus system  \cite{gaubert1998duality}.  This representation, however, is not unique, so does not always converge to the correct cycletime \cite{ELPthesis}.

Critical circuits have been a staple in the area of tropical mathematics.  In this paper, we have taken this a step further by adding edges to the circuit to connect the whole network as a reduced network.  The reduced network echoes recent work by Correia et al. who talk of redundancy and a ``metric backbone" of a complex network \cite{Rocha2018}.  Indeed, we have shown that, asymptotically, input that exceeds the threshold $\omega$ becomes redundant such that each node possesses the reduced neighborhood of $m=\lceil \omega|\mathcal{N}_i|\rceil$ affecting nodes.  In turn, our reduced network for some $\omega$ is equivalent to the aforementioned backbone.   We can go further and propose that if circuits and edges are common in attractor networks across many values of $\omega$, then they can be regarded as a `universal backbone' of the network.  We also note the feature of circuits in our backbone closely aligns with the work of Radicchi et al. \cite{radicchi2011information} in that we take a similarly higher-order approach to filtering: we don't simply remove the edges with smallest weight, but look at circuits.

\bibliographystyle{plain}
\bibliography{MaxminThreshold_arxiv}

\end{document}